\documentclass{aa}
\usepackage{txfonts}
\usepackage{natbib}\bibstyle{aa}
\usepackage{graphicx}

\newcommand{\figps}[2]{\resizebox{#1}{!}{\rotatebox{0}{\includegraphics{#2}}}}

\newcommand{\filps}[2]{\resizebox{#1}{!}{\rotatebox{-90}{\includegraphics{#2}}}}

\def\llm{{\sc LLmodels}}

\def\atlas{{\sc atlas9}}
\def\tef{T_{\rm eff}}
\def\tauros{\tau_{\rm ross}}

\def\kms{km\,s$^{-1}$}
\def\vturb{\upsilon_{\rm turb}}
\def\uvbyb{u\upsilon by\beta}

\begin{document}

\title{Theoretical analysis of the atmospheres of CP stars}
\subtitle{Effects of the individual abundance patterns}

\author{S. A. Khan\inst{1,2} \and D. V. Shulyak\inst{2}}

\offprints{S. A. Khan, \email{skhan@astro.uwo.ca}}

\institute{%
Physics and Astronomy Department, University of Western Ontario, London, ON, N6A 3K7, Canada \and
Institut f\"ur Astronomie, Universit\"at Wien, T\"urkenschanzstra{\ss}e 17, 1180 Wien, Austria }

\date{Received 30 January 2007 / Accepted 22 April 2007}
\abstract
{Historically, stellar model atmospheres with scaled solar abundances (for all elements heavier then helium) have been widely used for analysis of the atmospheres of chemically peculiar (CP) stars. However, in reality, atmospheres of CP stars demonstrate a variety of abundances, not necessarily scaled to the solar composition.}
{We study the effects of individual abundance patterns on the model atmospheres of CP stars. The main purpose is to conduct a systematic homogenous study to explore the abundance parameter space occupied by these stars.}
{We calculated a grid of the model atmospheres of A and B stars (${\log g=4.0}$) for different effective temperatures (${\tef=8000}$, 9500, 11\,000, 13\,000, 15\,000, 20\,000\,K) and chemical compositions. We used the \llm\ code to compute model atmospheres with individual abundance patterns, varying the following elements: C, Mg, Si, Ca, Ti, Cr, Mn, Fe, Ni, Sr, Eu and He. We compared the computational results for these peculiar model atmospheres with those of reference model atmospheres of the solar chemical composition.}
{We present a homogeneous study of model atmosphere temperature structure, energy distribution, photometric indices in the $\uvbyb$ and $\Delta a$ systems, hydrogen line profiles, and the abundance determination procedure as it applies to CP stars. In particular, we found that Si, Cr and Fe are the main elements to influence model atmospheres of CP stars, and thus to be considered in order to assess the adequacy of model atmospheres with scaled solar abundances in application to CP stars. We provide a theoretical explanation of the robust property of the $\Delta a$ photometric system to recognize CP stars with peculiar Fe content. Also, the results of our numerical tests using model atmospheres with one or several elements overabundant (Si and Fe by $+1$\,dex, Cr by $+2$\,dex) suggest that the uncertainty of abundance analysis in the atmospheres of CP stars using models with scaled abundances is less than $\pm 0.25$\,dex. If the same homogeneous models are used for the abundance stratification analysis then we find that the uncertainty of the value of the vertical abundance gradient is within an 0.4\,dex error bar.}
{Model atmospheres with individual abundance patterns should be used in order to match the actual anomalies of CP stars and minimize analysis errors.}
\maketitle
\section{Introduction} \label{introduction}

A fraction of at least 25\% \citep{schneider} of the upper main sequence stars is known as spectroscopically peculiar stars. Their spectra show anomalously strong (or weak) absorption lines of some chemical elements in comparison to those of normal stars with the same fundamental parameters. These stars are often called chemically peculiar (CP) stars, implying that unusual chemical composition (element abundances) rather than other causes is responsible for the spectrum anomalies. Consequently, this means that abundances in the atmospheres of CP stars are substantially different (enhanced or depleted) compared to the solar ones.

Effective temperatures of CP stars range from 6500\,K to about 25\,000\,K, corresponding to spectral types from late F to very early B. The widely used simplified classification of the CP stars by \citet{preston} includes (in order of increasing effective temperature): CP1 (Am/Fm stars), CP2 (Si, Cr-Sr-Eu classic magnetic A and B stars), CP3 (Hg-Mn, B-type stars) and CP4 (He-weak, B-type stars). The nomenclature specified in brackets reflects which elements are responsible for the principal visible peculiarities (usually enhanced) on low-dispersion spectrograms. For Am/Fm, these elements are Fe-peak and heavier elements. Note that there are no gaps in degree of peculiarity or in temperature limits between these subgroups themselves or with respect to normal stars. In fact, their domains smoothly overlap each other, often providing temperature dependant spectra anomalies (abundance effects); however this does not mean that they are related in terms of the origin of their peculiarities.

Subsequently, photometric data and high resolution spectra analysis in the visual and ultraviolet regions confirmed primary abundance peculiarities and revealed new features of CP stars. As a result, modern conventional classification is based on the presence of a magnetic field \citep{wolff_a, borra_landstreet, shorlin} and leads to two main subgroups of CP stars. The sequence of magnetic CP stars (mCP) from the lowest temperature is Ap and Bp (Eu-Sr-Cr-Si), He-weak (Sr-Ti, Si) and He-rich B-type stars. The sequence of non-magnetic stars (in the same $\tef$ order) is Am, Ap Hg-Mn, and He-weak (P-Ga) Bp stars. The field strength observed in magnetic stars ranges from a few hundred gauss to several tens of kilogauss. Other small subgroups of peculiar upper main sequence stars (e.g. $\lambda$~Bootis stars) usually are not considered within the framework of this classification.

\subsection{Observable abundances of CP stars} \label{introduction1}

\medskip\noindent\textsf{\emph{Am stars}.}
The classic definition for Am stars by \citet{conti} states that the Am class stars show apparent underabundant Ca (usually measured by the \ion{Ca}{ii} $\lambda$ 3933 (K) line) and/or an overabundance of the Fe group and heavier elements. The Ca deficiency may or may not be accompanied by underabundant Sc. Typical abundance values for elements in comparison to those in the solar atmosphere are from $-0.5$ to $-1.0$\,dex for C, Ca, Mg and Sc, and from $+0.5$ up to $+1$\,dex for the iron-peak (as Ti, Cr, Mn, Fe) and heavier elements. This includes the rare earth elements (REE) which are enhanced by up to $+1$\,dex as well. The range of effective temperatures for this class is 7000--10\,000\,K.

\medskip\noindent\textsf{\emph{Ap stars}.}
The spectra of Ap stars show enhanced lines of Si (up to $+2$\,dex), Sr and Cr (up to $+3$\,dex). Many stars show an excess of the iron-peak elements from $+1$ to $+2$\,dex. The REE elements (with an accent on Eu) are typically greatly increased and their abundance values range from $+3$ up to $+6$\,dex \citep[see for example][]{bonsack_wolff, strasser, bagnulo}. The C, N and O elements are typically underabundant by about $-1$\,dex \citep{roby_lambert}. The deficiency of He is found to be more substantial for Ap than Am stars and spans from $-1$ to $-2$\,dex. For Ap stars the abundances of the Fe-peak elements are not obviously correlated with the REE anomalies. Effective temperatures range from 6500 to 16\,000\,K.

\medskip\noindent\textsf{\emph{Hg-Mn stars}.}
The group of Hg-Mn stars shows overadundant Hg from $+3$ to $+6$\,dex \citep{jomaron} and Mn that can reach up to $+3$\,dex \citep{dolk}. Their spectra can exhibit overabundance of some other elements (such as P, Ga up to $+2$\,dex, and others) and deficiency in Al, Zn, Co and others. The Hg-Mn stars are also peculiar in He which is deficient by as much as $-1.5$\,dex \citep{dworetsky}. These stars occupy a range of effective temperatures from 10\,000 to 15\,000\,K.

\medskip\noindent\textsf{\emph{He-weak stars}.}
The He-weak stars have much weaker helium lines than those in spectra of stars with the same photometric colors and hydrogen line profiles. The helium deficiency spans from about factor of 2 (0.3\,dex) to 15 ($\sim$1.2\,dex). The group of He-weak stars consists of three subgroups: the P-Ga subgroup continues the HgMn stars towards higher temperatures; the Si and Sr-Ti subgroups are hotter analogs of Ap stars adjoining to the He-rich stars. In that sense, the He-weak group approximately continues the peculiar element abundances of its cooler and hotter analogs (for example strong Cr lines). Typical values of $\tef$ range from 12\,000 to 18\,000\,K.

\medskip\noindent\textsf{\emph{He-rich stars}.}
Finally, the group of He-rich (or He-strong) stars, which are main sequence stars, show abnormally high helium abundance (${N_{\rm He}\approx N_{\rm H}}$), while their effective temperatures are 21\,000--25\,000\,K. This group is outside the scope of this work due to its extreme nature and will not be considered in the current study.

The general values for chemical composition given above turn out to be even more complicated in reality because ``to those who really familiar with the spectra, some stars \emph{within} a given class are \emph{as} different from one another as they are from stars in a different class" \citep[][about Ap stars]{cowley} and ``large abundance differences between stars of similar temperatures and peculiarity classes may occur and do not correlate in any obvious way with other physical properties" \citep[][about CP stars]{bonsack_wolff}. That is why we will use observed abundances in our study as suggested reference values only.

\subsection{Spectra analysis and model atmospheres} \label{introduction2}

CP stars, on the average, are much slower rotators (${\upsilon_e\sin i\lesssim 50}$\,\kms) than normal stars (except for the He-rich stars), which makes possible \emph{detailed} line profile studies, providing better understanding of the CP stars phenomena. Spectra of magnetic stars show photometric, luminosity  and line profiles variations (accompanied by the magnetic field variability with the same period), whereas those of non-magnetic stars do not \citep[although some exceptions apply, see for example][about Hg-Mn stars]{adelman_hgmn}.

The explanation for such variations is well developed, and is known as the rigid (oblique) rotator model (where oblique refers to the magnetic axis, which is inclined with respect to the rotation axis) with a non-uniform distribution of chemical elements over the stellar surface. Moreover, many magnetic and perhaps non-magnetic CP stars show signs of non-uniform distribution of the chemical elements with depth (abundance stratification) in their atmospheres \citep[see e.g.][]{wade, savanov}.

Here we want to draw attention to the well known fact that model atmospheres are required to determine stellar atmosphere abundances, which in turn are essential input parameters for stellar evolution theory (stellar structure modelling). For example, recently revised solar photospheric abundances \citep{asplund} due to development of a new highly sophisticated solar model atmosphere influenced theoretical helioseismological predictions in a major way \citep[see e.g.][]{guzik,antia}. This illustrates why model atmospheres are a crucial part of the stellar astrophysics.

In the case of CP stars, the surface non-uniformity, vertical stratification, presence of the magnetic field, and the excess line blanketing make the spectral analysis quite complicated and uncertain. The usual analysis assuming a homogenous atmosphere models with scaled abundances and without magnetic fields \citep[e.g., the famous classic \atlas\ models by][]{kurucz13} may result in incorrect conclusions if such effects actually take place but are neglected in stellar atmosphere modelling.

Despite this concern, classic model atmospheres are widely used for a variety of stellar astrophysics problems. They have numerous applications to CP stars: determination of fundamental parameters, abundance analysis, stellar magnetic field geometry research, detailed line profile study (including full treatment of the Zeeman effect), and reconstruction of stellar surface properties by the Doppler Imaging (DI) technique (including all Stokes parameters).

The problem is that while the correct spectrum analysis procedure is an \emph{iterative} process of comparison between observational data, theoretical prediction, and the following corrections to the input modelling parameters, historically, it has developed that construction of model atmospheres and spectrum synthesis are almost completely separated procedures. The study done by \citet{lester} is one of the rare attempts to analyse the effects of this separation. He found a small systematic error affecting the determination of stellar abundances if the line opacity background is neglected during the spectrum synthesis procedure, while the same opacity is fully considered during model atmosphere calculation.

In fact, currently almost all spectral line profile analysis (including Stokes analysis) and abundance determinations are performed on the basis of classic model atmospheres (like \atlas\ models), without taking into account magnetic field effects, possible vertical abundance stratification, or even the specific chemical composition other than simply scaled to the solar abundance table. An initially picked model atmosphere is used for the whole analysis, usually not being recalculated in accordance with results of this analysis. An indicative note by \citet{adelman1995} says that we need model atmospheres ``that more closely match the actual abundance anomalies" to understand observable phenomena of CP stars.

However, it should be noted that several authors have constructed and used individualized/peculiar models in their research. For instance, \citet{kupka2} used a technique of model atmospheres calculation developed by \citet{piskunov_kupka} (who applied the Opacity Distribution Function method to CP stars) to investigate the famous depression around 5200\,\AA\ in spectra of CP stars. \citet{llmodels} used a model atmosphere with individual and stratified abundances to fit the 5200\,\AA\ depression, energy distribution and hydrogen line profile simultaneously for a hot CP star. \citet{leblanc_poprad} made a big step forward calculating models with stratified abundances in a self-consistent way. The most recent papers by \citet{paper2, paper3} revealed effects of the magnetic line blanketing due to the Zeeman splitting and polarized radiative transfer on the model atmosphere structure and observable characteristics of mCP stars.

Moreover, recent \emph{high resolution} stellar spectra analyses \citep{wade2006, lehmann, stutz1, fossati, silvester, nicole} as well as photometric analyses \citep{stutz2} of CP stars done with help of the \llm\ code show that it is practical for model atmospheres with individual abundance patterns to be routinely utilized for spectral study.

The main purpose of this paper is to study basic effects of the individual abundance patterns on the model atmosphere structure, observable characteristics and abundance analysis of the non-magnetic CP stars. While it is obvious that all features of CP stars should be included in a whole cycle of the stellar spectra modelling, no systematic quantitative homogenous research has been attempted on this question before.

This study continues our long-term investigation of model atmospheres of CP stars, highlighting some particular problems that observers encounter. Because of the complexity of the theoretical phenomena and difficulties of working with real spectra, we focus on one of many problems associated with CP stars at a time. We provide indicative results that may help to estimate how uncertain specific aspects of spectral analysis can be.

\section{Calculation of model atmospheres} \label{calculation}

Model atmospheres with individual abundance patterns calculated in this study were constructed by means of the latest version of the \llm\ code (version 8.4) which uses direct treatment of the line opacity \citep{llmodels} and provides us with a powerful tool to calculate model atmospheres of both CP and mCP stars within the LTE (local thermodynamical equilibrium) approach. The code has been successfully applied for a number of studies (see references in the previous section).

\subsection{Technique} \label{technique}

To cover the stellar parameter space appropriate for CP stars, we calculated a grid of model atmospheres with effective temperatures ${\tef=8000}$, 9500, 11\,000, 13\,000, 15\,000, 20\,000\,K, a single but typical value of the surface gravity ${\log g=4.0}$ and for a variety of abundance values.

All models were calculated for the whole range of effective temperatures even if some of them are outside the actual temperatures associated with a given class of CP stars (see Sect.~\ref{introduction1}). For example, models with overabundant Fe and Cr were calculated for $\tef$ up to 20\,000\,K. This way we cover the whole parameter space with model atmospheres to get a complete picture of changes and to analyse for any interesting behaviour.

As usual, a standard reference point to specify a peculiar chemical composition was set to the solar abundances denoted hereafter as ${\rm [M/H]=0.0}$ (where ${{\rm [X/H]\equiv[X]}=\log(N_{\rm X}/N_{\rm H})-\log(N_{\rm X}/N_{\rm H})_{\mbox\sun}}$ and $\rm M$ means metals, i.e. any chemical element except H and He). Also, if the designation in square brackets is used then we omit the dex units. The list of the adopted peculiar abundance patterns is represented in the next section (Sect.~\ref{adopted_abund}).

We used VALD \citep{vald1,vald2} as a source of spectral line data, including lines that originate from predicted levels. In total we extracted about 22 million lines; we then performed a preselection procedure to eliminate those that do not contribute significantly to the line opacity.

For the line preselection procedure we used model atmospheres calculated by \atlas\ \citep{kurucz13} with corresponding fundamental parameters from our grid using the Opacity Distribution Function (ODF) tables for the  largest value of the metallicity ${\rm [M/H]=+1.0}$ available on the Kurucz CD \citep{kurucz14}. At the same time we adopted the overabundant chemical composition ${\rm [M/H]=+2.0}$ to insure selection of as many spectral lines as possible. The preselection was performed once for each set of models with the same effective temperature (but various abundance patterns). The line selection criterion, which is the relation between the line and continuum absorption coefficients at the center of each line, was set to 0.1\% (ten times less than we usually use). For the number of spectral lines preselected, and thus involved in computing the blanketing, refer to Table~\ref{number_lines}.

\begin{table}
\centering
\caption{Number of spectral lines ($N_{\rm lines}$), wavelength intervals ($\lambda_{\rm range}$) and corresponding numbers of spectral wavelength points ($N_{\lambda\,\rm{points}}$) for different effective temperatures ($\tef$)  used for the opacity calculation. Lines were selected with a selection criterion 0.1\% using Kurucz's \atlas\ model atmospheres \citep{kurucz13}, ODF tables for ${\rm [M/H]=+1}$ and enhanced chemical composition ${\rm [M/H]=+2}$ to insure conservative preselection.}
\begin{tabular}{rccc}
\hline\hline
$\tef$, K  & $\lambda_{\rm range}$, \AA & $N_{\lambda\,{\rm points}}$ & $N_{\rm lines}$ \\
\hline
   8000    & 500--50\,000 & 495\,000 & 2\,693\,251 \\
   9500    & 500--50\,000 & 495\,000 & 2\,760\,456 \\
11\,000    & 500--30\,000 & 295\,000 & 2\,870\,658 \\
13\,000    & 100--30\,000 & 299\,000 & 3\,267\,218 \\
15\,000    & 100--30\,000 & 299\,000 & 3\,342\,524 \\
20\,000    & 100--30\,000 & 299\,000 & 3\,462\,336 \\
\hline
\end{tabular}
\label{number_lines}
\end{table}

To calculate model atmospheres of our grid, we used either \atlas\ model atmospheres with metallicities ${\rm [M/H]=+1.0}$, or model atmospheres already calculated within this study (whichever was more suitable) as an initial guess of the model structure to iterate from. Primarily to obtain faster convergence, we tried to use models with a close abundance pattern to continue from them to more peculiar model atmospheres. Some models for ${\tef=20\,000}$\,K required use of the grey approximation to make an initial model atmosphere structure, to avoid temperature correction freezing in the upper atmosphere.

\begin{table*}
\centering
\caption{The table represents different abundances adopted for model atmosphere calculations in this study. All abundances are given on a logarithmic scale with respect to the solar composition (i.e. dex units). Some elements and their respective values are accompanied by other element-value pairs indicated in parentheses. This designation means that models were calculated not only for abundance values of the element specified in its column but also for combinations of these elements with abundance variations of other elements specified in parentheses. If ``all" is specified then all values listed in the respective column are combined with the main element value outside the parentheses. The star symbol ($\star$) marks the reference solar composition model, the asterisk symbol ($\ast$) indicates non-typical abundance values for CP stars (see Sect.~\ref{special_changes}), and the diamond symbol ($\diamond$) points to the models with a decreased amount of He in order to normalize the sum of all abundances.}
\begin{tabular}{lccclllcllllcccc}
\hline\hline
$\rm [M/H]$ & He & C    & CNO  & Mg   & Si\,(Fe or Cr) & Ca   & Ti   & Cr\,(Fe) & Mn\,(Hg) & Fe  & FeSiCr & Ni   & Sr   & Eu   & Hg   \\
\hline
$\,\,\,\,0.0\star$ & $-0.417$ & $-1$ & $-1$ & $-1$ & $+1\,({\rm all})$ & $-1$ & $+2$ & $+1\,({\rm all})$ & $+1$ & $+1$ & $+1$ & $+2$ & $+2$ & $+2$ & $+5$ \\
$+1.0$ & $-0.894$ & $-2$ &      &$+2*$& $+2\,({\rm all})$ &$+2*$&      & $+2\,({\rm all})$ & $+2$ & $+2$ & $+2$  &   & $+3$ & $+5$ &      \\
$+2.0\diamond$ & $-1.894$ &      &      &      & $+3\,({\rm all})\diamond$ &      &      & $+3\,({\rm all})$ & $+3\,(+5)$ &      &   &      &      &     &  \\
\hline
\end{tabular}
\label{abundances}
\end{table*}

A logarithmic Rosseland optical depth scale ${\log\tauros}$ was used as an independent variable of atmospheric depth spanning from $+2$ to $-6.875$ and subdivided into 72 layers. Convection and magnetic fields were neglected and a zero value for the microturbulent velocity $\vturb$ was adopted to exclude their possible influence and focus on the abundance pattern effects.

\subsection{Adopted abundances} \label{adopted_abund}

Different chemical compositions were chosen in accordance with the typical observable abundance patterns found in spectra of CP stars (see Sect.~\ref{introduction1}). Consequently, we adopted various abundance values for the following elements: He, C, N, O, Mg, Si, Ca, Ti, Cr, Mn, Fe, Ni, Sr, Eu and Hg. Note that it is obviously almost impossible to perform a clear analysis on the basis of a large variety of abundance combinations. Thus, we chose to analyse the influence of each element separately, as a preliminary stage of the study, to understand which elements deserve more attention. Besides typical amounts of the elements attributed to CP stars we considered several abundance values outside of those typical limits (see Sect.~\ref{special_changes} for a description) as well as scaled to solar chemical compositions in order to extend our study.

The detailed data about all chemical elements considered and their respective values assumed in this study are presented in Table~\ref{abundances}. Some model atmospheres were calculated for combinations of several chemical elements (which have been found the most influential elements). Such cases are specified both in the header of the table and in the table itself in the parentheses. 

For He-weak models we used three values of the depletion ${N_{\rm He}/N_{\rm total}=0.001}$ ($-1.894$\,dex), 0.01 ($-0.894$\,dex), 0.03 ($-0.417$\,dex), while the solar amount is equal to 0.0783 (0.0\,dex). The hydrogen abundance was accordingly modified (increased) to normalize the sum of all abundances. The $M/N_{\rm total}$ value for the rest of the abundances stays the same as in the solar atmosphere (though in the ${\log(M/N_{\rm H})}$ scale they are decreased by 0.022--0.035\,dex).

Model atmospheres with scaled solar abundances were calculated for the values of ${\rm [M/H]=+1.0}$, $+2.0$ and for the solar composition (${\rm [M/H]=0.0}$). The latter models are called hereafter \emph{reference model atmospheres} because they are used to perform the comparisons made in this study.

The actual solar amount of each element adopted in this study was taken from the recent work by \citet[][Table~1]{asplund} and converted into the $N/N_{\rm total}$ scale, where the fraction of hydrogen is 0.921.

Some model atmospheres required a decreased amount of He so the sum of all abundances would be normalized (see elements in Table~\ref{abundances} marked with a diamond ($\diamond$) symbol). This includes model atmospheres with abundances scaled up by $+2$\,dex (He is 0.002) and model atmospheres with Si overabundant by $+3$\,dex (He is 0.0483). For the rest of peculiar model atmospheres the sum of all abundances differs from unity by less than 1\% (i.e. the sum of all elements ${\sum N_i=1.00\pm 0.01}$), thus we left hydrogen and helium abundances for them unchanged.

Finally, note that while for model atmospheres peculiar in some chemical elements the results are pretty obvious in advance, nevertheless we adopted the peculiar values of those elements and performed the whole set of calculations in order to provide clear evidence for this point within the framework of the used data and techniques.

\section{Numerical Results}\label{results}

In this section we present the numerical results of the modelling: the effects of the individual abundance patterns on the model atmosphere temperature structure, energy distribution, photometric indices, hydrogen line profiles, and on the abundance determination procedure.

The results are described with respect to the reference model atmospheres. This way we can clearly identify peculiar features provided by various abundance values of different chemical elements.

To find a generalized characteristic of which element produces larger changes (is more influential) we use measurements of the average and maximum deviations in the whole parameter space of effective temperatures and abundances.

While for a particular chemical element, usually the largest (or the smallest) abundance value produces the maximum effect, the comparison among different elements is not so straightforward. The main problem, making the comparison more complex, is that different chemical elements may have different abundance ranges assigned to them. As a result, an element which is found to be more influential than another one may in fact produce less effect than the latter for some abundances or effective temperatures.

It should be also remembered that there are some non-typical abundance values for some elements (marked by ($\ast$) in Table~\ref{abundances}) which are treated separately (Sect.~\ref{special_changes}) before drawing a conclusion about the influence of these elements.

\subsection{Model atmosphere temperature structure} \label{temperature}

In order to understand the influence of different abundances and their combinations on the model atmosphere structure, and thus on the abundance determination through analysis of the line profiles of particular absorption lines, we investigate the temperature structure differences between model atmospheres with peculiar abundances and those with the solar composition.

To represent such differences we find that the per cent scale (${\Delta T/T}$) provides a better view of the values of the changes than the absolute one. At the same time, for better understanding, it is convenient to know the absolute value rather than relative values. We note that a 1\% change corresponds approximately to 50--90\,K for ${\tef=8000}$\,K or 100--200\,K for ${\tef=20\,000}$\,K depending on the optical depth (from surface layers down to $\log\tauros=0$). While this approximation is rather rough and depends on the effective temperature and particular shape of the temperature profile, we find that it is quite useful.

We decided to divide all chemical elements into three groups according to the level of changes in the temperature structure they show in comparison to solar composition models with the same fundamental parameters (the reference models). A given element falls into a certain group if all the model atmospheres solely peculiar in this element demonstrate temperature structure changes which do not overstep the limits assigned to that group.

\subsubsection{Small ($\leq$1\%) changes: He, C, N, O, Mg, Ca, Sr, Eu, Hg} \label{small_changes}

In this group we consider chemical elements those influence on the model temperature structure is not more than $\pm 1$\% in comparison to the corresponding reference models with solar chemical composition.

Models of the He-deficient stars demonstrate about 0.2\% (or 20\,K) temperature decreasing on the average and this cooling does not exceed ${0.5}$\% (or 60\,K) for the highest value (20\,000\,K) of the effective temperature. For ${\tef\leq 13\,000}$\,K such changes are noticeable only from the optical depth ${\log\tauros=-2}$ upwards the atmosphere. The temperature difference between models with the highest and the lowest He deficient values (${N_{\rm He}/N_{\rm total}=0.001}$ and 0.03) grows a bit with increasing $\tef$ but does not become larger than 0.1\% (or 10--15\,K) for the highest value. Here we note that analogous results were obtained in the study by \citet[][Sect.~3]{glagol_he} where He-deficient model atmospheres were calculated by D.~Shulyak with help of the \llm\ code in a similar way to the present study. We found that the variation of the mean molecular weight due to He deficiency is partly responsible for these small changes in the model atmosphere structure (see Sect.~\ref{small_flux}).

For model atmospheres with the deficient element C, temperature changes do appear for high effective temperatures (${\tef>11\,000}$\,K), and show an approximately constant shift (plateau) with respect to the solar model, with the temperature about 1\% (or 80--100\,K) lower above ${\log\tauros=0}$. For low effective temperatures (${\tef\leq 11\,000}$\,K) the maximum deviation is only as much as 0.3\% (or 30\,K) in some layers. The difference in the temperature structure between models deficient in C by $-2$ or $-1$\,dex is less than 0.2\% (or 20\,K), with more deficient models showing lower temperatures.

The combined influence of the elements C, N and O each decreased by 1\,dex results in a very small effect as well. The temperature structure falls in between the structures of the C-deficient models ($-1$ and $-2$\,dex) and its deviation from the solar composition model grows from a vanishing value for the lowest effective temperature up to 0.2\% (or $\sim$15--20\,K) for the highest ${\tef=20\,000}$\,K.

Negligible changes of about 0.05\% (or 5\,K) in the range of the optical depths from ${\log\tauros=-3}$ downwards the atmosphere are demonstrated by the models underabundant by 1\,dex in the elements Mg or Ca. For the upper atmosphere layers (${\log\tauros<-3}$) such changes are larger and amount to at maximum 0.1--1\%.

The model atmospheres with overabundant Sr (+2, +3\,dex) and Eu (+2, +5\,dex) differ from the solar composition models only by about 0.2\%, which is clearly within the range of the computational errors. Actually, the temperature structure of the models overabundant only by $+2$\,dex in Sr or Eu is practically indistinguishable from that of the solar composition models. The weak influence also can partly be explained by the lack of data about lines of the REE.

Finally, an enormously increased amount of Hg ($+5$\,dex) produces almost no changes in the model atmosphere temperature structure. In fact, it has the smallest influence among all elements considered above: specifically, it gives a 0.01\% (or 1--2\,K) change.

\begin{figure*}
\filps{\hsize}{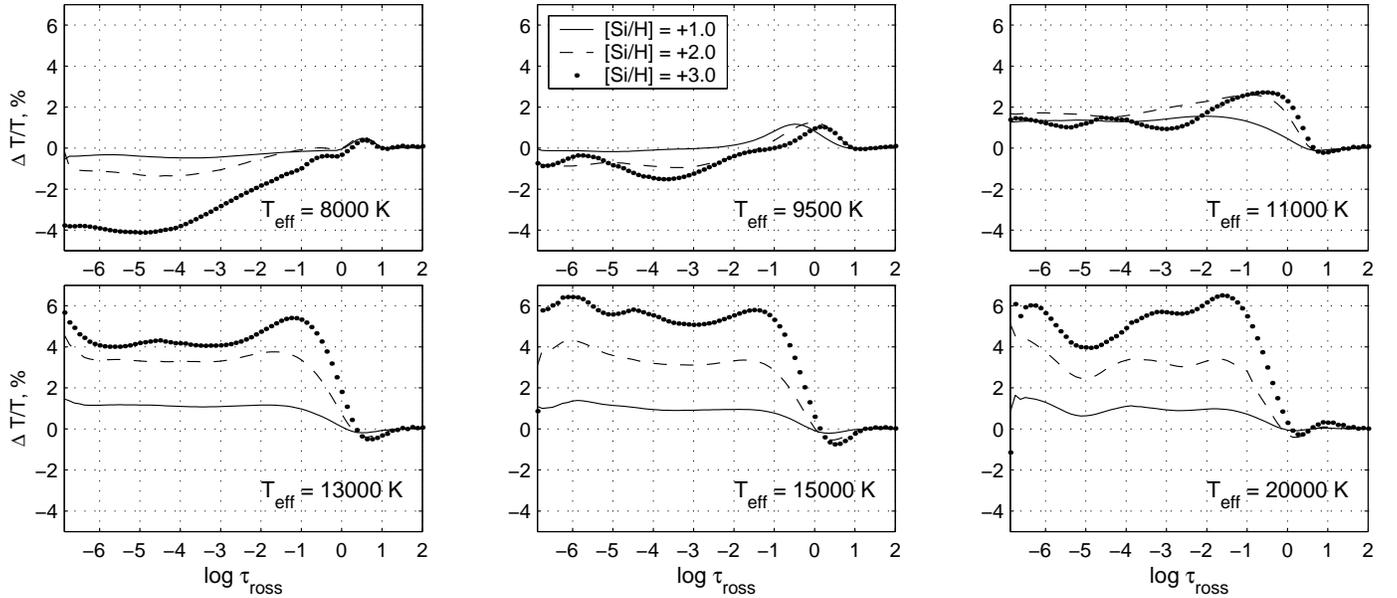}
\caption{Temperature differences between model atmospheres overabundant in Si by $+1$, $+2$ and $+3$\,dex and the solar composition models for the set of effective temperatures. The differences are represented on the per cent scale ${\Delta T/T}$ with respect to the solar composition models. The optical depth scale $\log\tauros$ is a logarithmic Rosseland scale.}
\label{si}
\end{figure*}

\subsubsection{Setting the error bar threshold} \label{error_bar}

Despite the fact that some changes in the temperature structure can be confidently attributed to calculation errors while some are real changes, we have set the 1\% level as a threshold limit and consider all deviations below this limit as insignificant and unreliable results. Though the same code was used throughout the whole work, we note that a variety of numerical problems (e.g. the transfer equation solution) have different valid methods to solve them, and once such methods are replaced with others, one would easily expect changes in the temperature structure of order 1\%. The accuracy of the line list data is also limited. Moreover, it seems highly unlikely that the observations can provide us with data accurate enough to perform any analysis with a precision higher than 1\%. However, these concerns do not affect the generally high internal precision of the study itself, since all the models have been calculated the same way with the same basic physics, atomic data, modelling assumptions, etc.

\subsubsection{Moderate (1--3\%) changes: Ti, Mn, Ni} \label{moderate_changes}

Model atmospheres and corresponding chemical elements assigned to this group produce changes in the temperature structure that clearly exceed the 1\% limit for some effective temperatures but are almost always below the 3\% limit. The elements Ti, Mn and Ni, which belong to this group, are Fe-peak elements and can be overabundant in stellar atmospheres (see Sect.~\ref{introduction1}).

For models with excess Ti and Ni, the main line forming region does not show changes that are actually larger than 0.3\% for Ti, and 0.35--1.2\% for Ni. Precisely, the region with less than $\sim$1\% changes spans from the deepest layers up to about ${\log\tauros=-3}$ optical depth. Changes of more than 1\% occur only above that level (exceeding 3\% for Ni in some layers). We found that the maximum changes in the temperature structure appear at 11\,000\,K for Ti, and at 20\,000\,K effective temperature for Ni.

\begin{figure*}
\filps{\hsize}{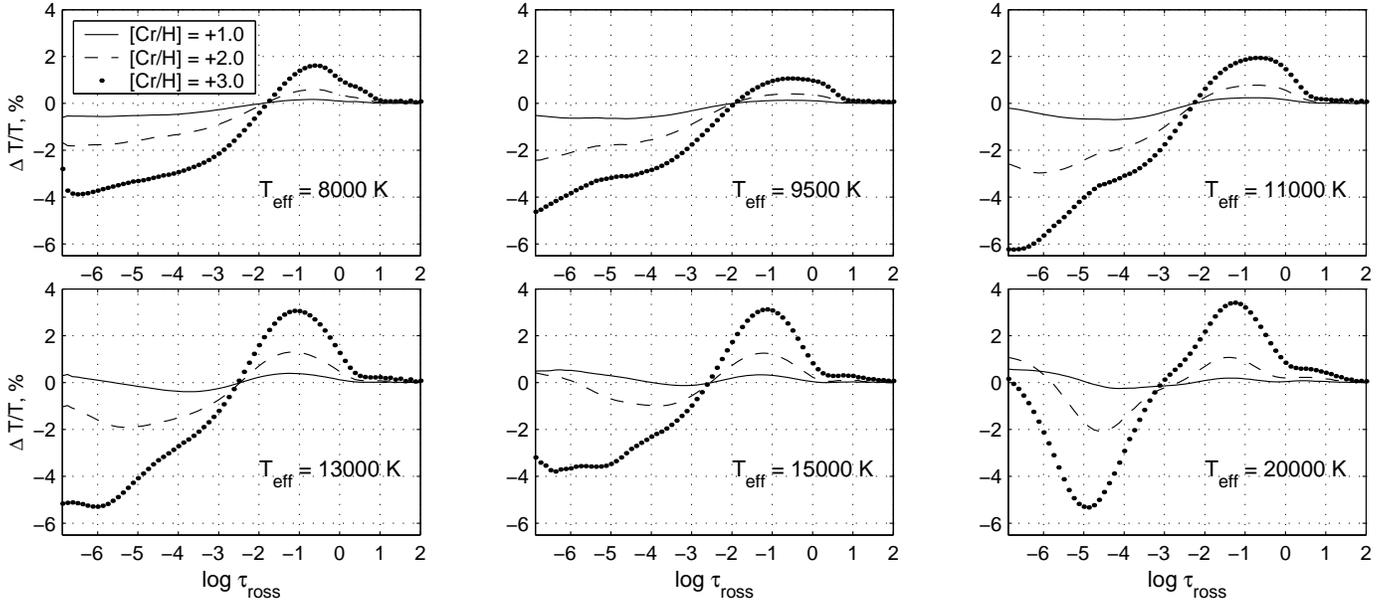}
\caption{Temperature differences between model atmospheres overabundant in Cr by $+1$, $+2$ and $+3$\,dex and the solar composition models for the set of effective temperatures. The axes are the same as in Fig.~\ref{si}.}
\label{cr}
\end{figure*}

The element Mn produces more profound effects on the model structure, increasing the temperature in the line forming region by about 1--2\% and steadily decreasing it by $\sim$2\% in the outer layers (${\log\tauros<-3}$). This temperature behaviour is very close to that demonstrated by models with enhanced Cr and Fe (see Sect.~\ref{large_changes}), where two distinct regions of heating and cooling determine the overall temperature profile. In particular, Mn produces changes which are closer in shape to those produced by Cr (see Fig.~\ref{cr} as an illustration) than those by Fe, although all of them generally are quite similar.

The combined influence of ${[\rm Mn]=+3}$ and ${[\rm Hg]=+5}$ on the model atmosphere structure does not differ from that of ${[\rm Mn]=+3}$ alone. The maximum deviation is about 0.1\% (or 10\,K) for the highest ${\tef=20\,000}$\,K, in the surface layers only. That is not surprising, since Hg belongs to the small changes group (Sect.~\ref{small_changes}).

\subsubsection{Special +2\,\,dex case: Mg, Ca} \label{special_changes}

In this section we consider model atmospheres with chemical compositions which are not generally associated with CP stars (see Sect.~\ref{introduction1}). The reason for this is that, in fact, different chemical elements are often stratified in the atmospheres of CP stars \citep{ryabchik_strat,ryabchik_2005} and the actual element content in different atmospheric layers may be different, not necessarily the one deduced from the classic analysis (i.e. some layers may show non-typical enhanced or depleted values).

At this point of our study, several chemical elements were assigned to the small changes group, according to the influence they have on the model atmosphere structure with the adopted abundance values. The fact is that some of these elements (Mg, Ca) are found to be stratified in the atmospheres of CP stars and can be strongly overabundant in some atmospheric layers. While it is clear that the elements not assigned to the small changes group (such as Si, Cr, Fe, etc.) are quite important in \emph{any} variations, the role of Mg and Ca if they are overabundant is unclear. So, we decided to consider these two elements overabundant by $+2$\,dex (somewhat overestimated value, see notes about Ca and Mg in Sect.~\ref{small_flux}, \ref{moderate_flux} respectively) to complete the analysis of the small changes group elements. We consider these abundance values as a special case and treat them separately from typical underabundant values. They are marked with the asterisks ($\ast$) in Table~\ref{abundances}.

Models with the excess of Ca can be assigned to the group of models with small changes in the temperature structure (Sect.~\ref{small_changes}) because only one model for ${\tef=11\,000}$\,K oversteps the 1\% limit for the very surface layers (${\log\tauros<-5.7}$) and reaches a 2\% negative deviation. For high effective temperatures (${\tef>11\,000}$\,K), the models with ${[{\rm Ca}]=+2}$ show a steady temperature increasing for the range of optical depths ${\log\tauros<-1}$ within the 0.5--1.0\% bounds.

The temperature changes for models overabundant in Mg fall into the group of the moderate effects (Sect.~\ref{moderate_changes}). For low effective temperatures (${\tef\leq 11\,000}$\,K) the changes are still within the 1\% threshold (except a narrow deviation by 1.6\% in between ${\log\tauros=-1}$ and $0$ for ${\tef=8000}$\,K) while they exceed this limit for a wide range of optical depths with ${\log\tauros<-1}$ by about 2\% (or 200\,K) for the higher temperatures.

\subsubsection{Large ($>$3\%) changes: Si, Cr, Fe and [M/H]} \label{large_changes}

In this section we consider the most interesting group of elements (including scaled abundances) which produce the most noticeable temperature variations (${\Delta T/T>3}$\%) in the model temperature structure.

\medskip\noindent\textsf{\emph{Silicon}.}
The comparison between temperature structures of the Si-peculiar and the solar composition model atmospheres are shown in Fig.~\ref{si}. The main distinctive feature of the Si temperature profile is a broad plateau of increased temperature from about ${\log\tauros=0}$ upwards throughout the whole atmosphere for high values of the effective temperature (${\tef\geq11\,000}$\,K). For low effective temperatures (${\tef<11\,000}$\,K), there are no very remarkable changes in the model structure, with the exception of the ${\tef=8000}$\,K, ${[{\rm Si}]=+3}$ model, where the temperature decreases considerably (even down to $-4$\%) for all optical depths ${\log\tauros<-1}$. We discuss the reason for this behaviour in Sect.~\ref{large_flux}.

The temperature deviations are about 0.5--1.5\% (or 50--100\,K) for models overabundant by $+1$\,dex, approximately 1--4\% (or 100--400\,K) for $+2$\,dex models, and about 1.5--6\% (or 100--800\,K) for models with Si enhanced by $+3$\,dex. As one can see in the figure, the ``abundance -- temperature deviation" relation has a large spread in values due to its extreme dependance on the effective temperature.

\medskip\noindent\textsf{\emph{Chromium}.}
Fig.~\ref{cr} represents the temperature differences between models with overabundant Cr and reference model atmospheres. All subplots in this figure show similar temperature profiles with temperature increasing deep in the atmosphere and cooling in the upper layers. This is the usual temperature profile that model atmospheres with enhanced opacities are expected to demonstrate, due to the backwarming and surface cooling effects. All three abundance values produce progressively increasing changes in the atmosphere structure, cooling the optical depth regions with reduced temperature and heating the regions with increased temperature.

The temperature deviation for the ${[{\rm Cr}]=+1}$ model atmospheres is about 0.5\% (or 50\,K), for models with ${[{\rm Cr}]=+2}$ its value is approximately 2\% (or 100--200\,K) and for model atmospheres peculiar in Cr by $+3$\,dex the changes are about 4\% (or 200--500\,K). The position (optical depth) of the inflection point separating the cooling and heating regions does not depend on the Cr abundance value and shifts evenly with the temperature increase from ${\log\tauros=-2}$ for ${\tef=8000}$\,K to ${\log\tauros=-3}$ for ${\tef=20\,000}$\,K. Looking ahead, we can say that Cr seems to demonstrate the most organized temperature behaviour among all elements.

\begin{figure*}
\filps{\hsize}{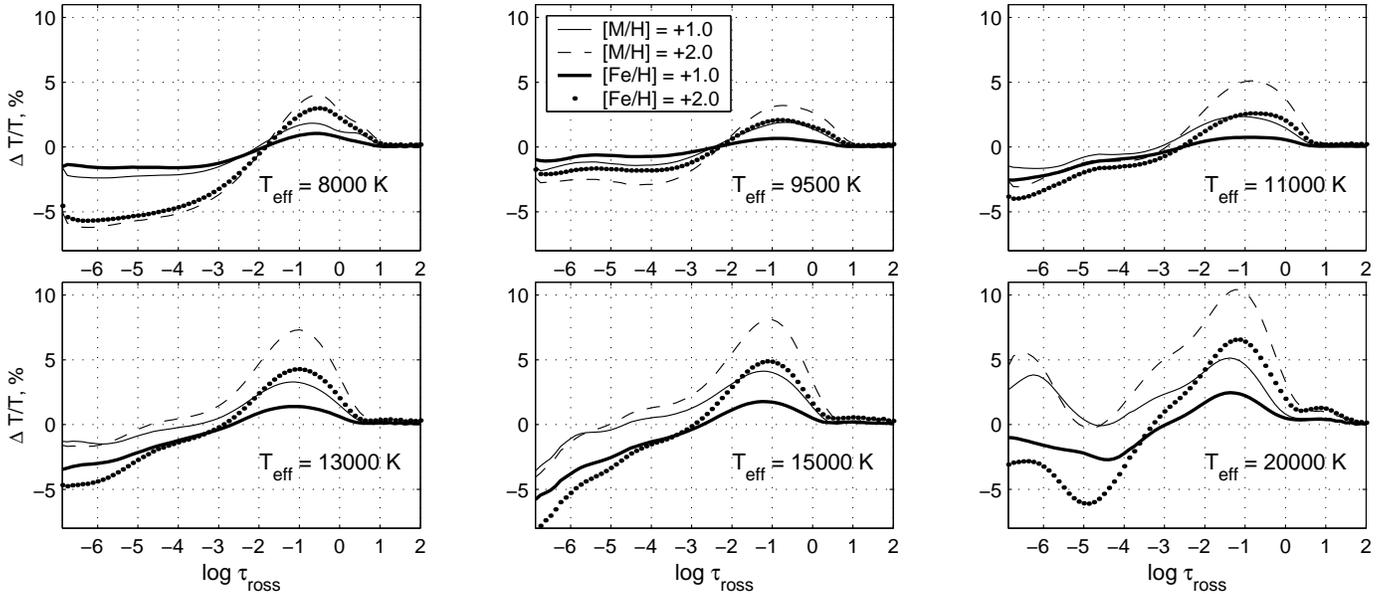}
\caption{Temperature differences for model atmospheres with scaled abundance and models overabundant in Fe by $+1$ and $+2$ for the set of effective temperatures. The axes are the same as in Fig.~\ref{si}.}
\label{a-fe}
\end{figure*}

\medskip\noindent\textsf{\emph{Iron}.}
The differences in the temperature structure between model atmospheres with overabundant Fe and the solar composition models are presented in Fig.~\ref{a-fe}. For comparison purposes, this figure also shows the temperature structure of the model atmospheres with scaled abundances. The influence of Fe is essentially the same as the influence of Cr: progressive heating and cooling of the deep and surface layers respectively as the Fe abundance increases. Generally, Fe demonstrates the most dramatic changes among elements chosen to be considered in this section. We will discuss the Fe influence more in the following sections.

The changes in the temperature structure for models overabundant in Fe by $+1$\,dex are about 1--3\% (or 100--200\,K) and for models with ${[{\rm Fe}]=+2}$ those changes are about 2--6\% (or 200--800\,K).

\medskip\noindent\textsf{\emph{Scaled abundances}.}
The temperature structures of model atmospheres with scaled solar abundances are also shown in Fig.~\ref{a-fe}. As one can see these models produce much stronger temperature effects in comparison to the purely Fe-peculiar models. Only for the lowest effective temperature ${\tef=8000}$\,K are models with scaled abundances similar to those Fe-peculiar models. The figure shows two evident heating and cooling regions in the model structure. The size of the heating region quickly grows, expanding towards the surface layers with increasing effective temperature. At the same time a weak cooling region shrinks more and more and even disappears for ${\tef=20\,000}$\,K.

Similar temperature patterns are shown by models with magnetic line blanketing \citep[][Fig.~1]{khan_paper1}, and by models with enhanced microturbulent velocity, most likely due to the fact that in all cases spectral lines of all chemical elements (rather than some of them) are strengthened. Another interesting detail is that ${\rm [M/H]=+2.0}$ models provide almost exactly twice as much heating as ${\rm [M/H]=+1.0}$ models do. Neither Fe nor Cr, Si or Mn show such a straightforward dependance, for any effective temperature.

Model atmospheres with abundances scaled by $+1$\,dex show changes in the temperature structure of about 2--5\% (or 150--800\,K), while for $+2$\,dex models these values are around 3--10\% (or 300--1600\,K).

\subsubsection{Discussion on the temperature structure} \label{changes_discussion}

Here we want to summarize, explore and discuss in detail the main results of the temperature structure analysis and draw some conclusions for Sect.~\ref{temperature}.

After the analysis, within the range of abundances selected for consideration in this study (and which are typical for CP stars) we found that the most important elements to influence the temperature structure of model atmospheres of peculiar stars are Si, Cr and Fe.

If we consider changes in the model atmosphere structure in the main line forming region (say from ${\log\tauros=-2}$ to 0) introduced by equally overabundant ($+1$ or $+2$\,dex) elements of Si, Cr and Fe taken separately, then, generally, the influence of Fe dominates over the influence of Si and Cr. The next most influential element is Si, which in some cases (for high $\tef$) produces temperature changes which even slightly exceed those due to Fe (${\tef=9500}$ and 11\,000\,K, $+1$\,dex) or are almost equal to them (e.g. ${\tef=13\,000}$\,K, $+2$\,dex). Comparing Si-enhanced models to models overabundant in Cr by $+3$\,dex shows that for low effective temperatures, Cr produces a larger (${\tef=8000}$\,K) or similar (${\tef=9500}$\,K) temperature deviation compared to Si, while for higher temperatures, Si overtakes Cr as was noted above. Consequently, putting iron, chromium and silicon in order of decreasing temperature influence in the main line forming region we obtain the sequence of Fe-Si-Cr, where Si tends to reach or sometimes even exceed the level of the Fe influence for high effective temperatures and sometimes yields to the influence of Cr for low effective temperatures. We find that this behaviour for Si is explained by its sharp temperature dependant influence profile noted above (Sect.~\ref{large_changes}). Having this element sequence in mind, later when we will consider their combinations in pairs we will test the influence of a ``weaker" element on a ``stronger" element (for example, how Si affects Fe-peculiar models or Cr affects Si-peculiar ones).

An interesting result is that some models demonstrate an organized temperature behaviour with two distinct heating and cooling regions in comparison to the reference models. Elements that produce such temperature changes are Cr, Mn and Fe (which actually appear following one another in the periodic table of the elements).

The heating region is located deep in the atmosphere and its magnitude clearly depends on the effective temperature and the element's overabundance value, steadily growing with the increasing of both of them for all mentioned elements. However, the temperature increasing pattern actually starts from ${\tef=9500}$\,K (not from the lowest temperature considered). This fact can be seen in either Fig.~\ref{cr} for Cr or Fig.~\ref{a-fe} for Fe, where the first subplots for ${\tef=8000}$\,K show stronger heating than the second subplots for ${\tef=9500}$\,K, while starting from the latter one, the temperature pattern seems to be held reliably. We did not provide a figure for Mn but it exhibits characteristics quite similar to those of Cr (including the dependence of the heating amplitude on the effective temperature). In addition to that, the scaled model atmospheres also demonstrate the same kind of behaviour, but even more pronounced (see Fig.~\ref{a-fe}) including the exception for the ${\tef=8000}$\,K value.

The cooling region shows only the abundance dependant temperature behaviour. The higher the abundance is, the cooler the upper atmosphere. Besides that there are no other evident common traits for all elements to point out.

Another feature for the group of Cr, Mn and Fe elements is that the inflection point which separates the cooling and heating regions progressively moves towards the surface with increasing effective temperature. For Mn and Cr, the position of this point does not depend on the abundance value but for Fe it does slightly and for high temperatures only (${\tef>11\,000}$\,K).

Model atmospheres with scaled solar abundances show temperature profiles that can not be simply replicated by overabundance of any single element separately (or otherwise). Even if some peculiar models are similar to those with the scaled abundances (e.g. models with ${[{\rm Fe/H}]=+2.0}$ and ${\rm [M/H]=+1.0}$ for ${\tef=9500}$\,K in Fig.~\ref{a-fe}), there is no way to find the best model substitution in advance, without actually calculating one and looking for a fit, if any. We confirm that model atmospheres with scaled abundances can not be used to simply simulate the temperature effects of the individual abundance patterns \citep[see][for a thorough discussion]{piskunov_kupka}.

We also tested the combined effect on the temperature structure of the most influential elements. We calculated model atmospheres overabundant in two elements simultaneously (e.g. Fe and Cr, Fe and Si) considering all possible combinations among the values used in this study (see Table~\ref{abundances}).

\begin{figure*}
\filps{\hsize}{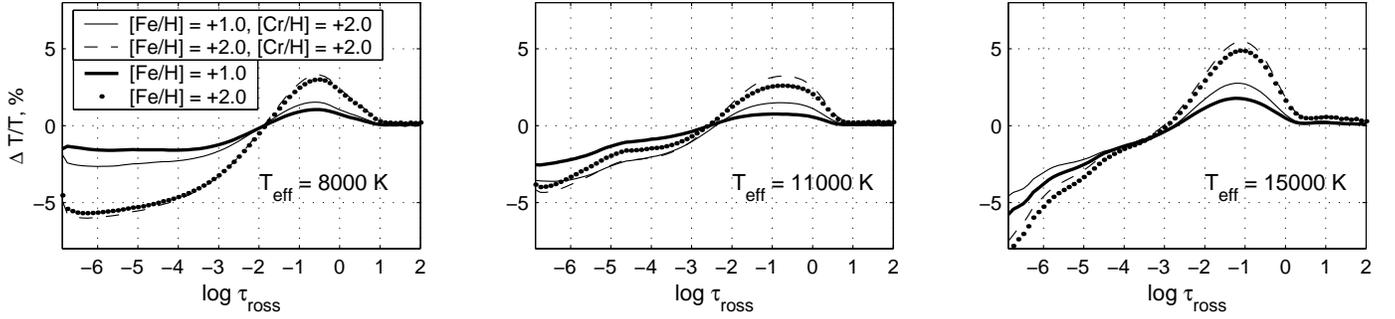}
\caption{Temperature differences of model atmospheres with individual Fe and Fe-Cr abundance patterns from the solar composition models. Subplots are shown for ${\tef=8000}$\,K, 11\,000\,K, 15\,000\,K. The figure demonstrates the weak influence of overabundant Cr in the Fe-enhanced model atmospheres. The axes are the same as in Fig.~\ref{si}.}
\label{fe-cr}
\end{figure*}

\medskip\noindent\textsf{\emph{Iron--chromium}.}
We found that Cr enhanced up to $+2$\,dex does not introduce much of a change in the temperature structure of the model atmospheres peculiar in Fe by $+1$ or $+2$\,dex. Figure~\ref{fe-cr} illustrates this statement. The low sensitivity of the Fe-peculiar models to enhanced Cr content is explained by the fact that ${[{\rm Cr}]\leq +2}$\,dex itself does not change the atmosphere structure a lot -- it produces less than a 2\% effect (see Fig.~\ref{cr}). What is even more important is that model atmospheres peculiar in Cr have temperature profiles similar to the Fe-peculiar models as was discussed in Sect.~\ref{large_changes}. Altogether this results in a weak influence of Cr enhanced by up to $+2$\,dex on the Fe-peculiar models, namely ${\lesssim 1}$\% (so the maximum influence is half of the Cr influence alone).

As shown above, models with Cr overabundant by $+3$\,dex show changes in the temperature structure which are quite large. When Cr overabundant that much is coupled with enhanced Fe there are obvious significant changes in the Fe-peculiar models due to Cr. However, it should not be forgotten that the ${[{\rm Fe}]=+3}$ value was not considered in this study since it is too high even for CP stars, thus comparison between models overabundant in Fe and in both Fe and Cr by $+3$\,dex was not possible. We conclude that if Cr is less or equally abundant than Fe then its influence can be considered as insignificant (below 1\%).

Generally, model atmospheres with a larger Fe content are less sensitive to the excess of Cr. For instance, in the atmosphere with solar Fe abundance value the maximum temperature deviation resulting from the influence of Cr overabundant by $+2$\,dex is 2\% on average (see Sect.~\ref{large_changes}), while the atmosphere peculiar in Fe by $+1$ ($+2$)\,dex is changed by ${[{\rm Cr}]=+2}$ by about 1.0 (0.5)\% respectively. In case a model is enhanced in Cr by $+3$\,dex then the maximum deviation produced due to Cr on average is about 4\% for the ${[{\rm Fe}]=0}$ model, 2\% for ${[{\rm Fe}]=+1}$ and 1.5\% for ${[{\rm Fe}]=+2}$.

\medskip\noindent\textsf{\emph{Iron--silicon}.}
Model atmospheres overabundant in both Fe and Si differ a lot from models enhanced only in Fe. The reason for this is the dramatically different effect of Si on the model atmosphere structure compared to that of Fe (compare Figs.~\ref{si} and \ref{cr}). Although the Fe-peculiar models additionally enhanced in Si by $+1$\,dex are found to show less than 1\% deviations with respect to those with no enhanced Si at all, the overall temperature changes which they show are very broad, covering almost the whole atmosphere. In case of ${[{\rm Si}]=+2}$ or ${[{\rm Si}]=+3}$ the contrast between structures of the purely Fe-peculiar and FeSi models is even more pronounced. Generally, an excess of Si seems to influence the temperature profile of the Fe-peculiar models in a unique way -- uniformly increasing the temperature from ${\log\tauros=0}$ up to the surface, as it does for purely Si-peculiar models (see Sect.~\ref{large_changes}).

The resulting effect of the exceptional influence of Si coupled with enhanced Fe is quite similar to what we see in the case of scaled solar abundances (compare the [Fe/H] and [M/H] profiles in Fig.~\ref{a-fe}). Indeed, our tests show that models peculiar in both Fe and Si by the same value ($+1$ or $+2$\,dex) resemble the temperature profiles shape of the model atmospheres with scaled abundances (${\rm [M/H]=+1}$ or $+2$) fairly closely for low effective temperatures (${\tef\leq 11\,000}$\,K). Of course, these FeSi models do not provide an exact fit but they are much closer than just Fe-peculiar models. If the Si abundance is increased more, by another $+1$\,dex, then the temperature profiles generally provide better fit to those models with scaled abundances with high effective temperatures (${\tef>11\,000}$\,K). For example, model atmospheres with ${[{\rm Fe}]=+2}$ and ${[{\rm Si}]=+3}$ exhibit a temperature structure that is very close (within ${<1}$\% error bar) to the ${\rm [M/H]=+2}$ model structure for ${\tef\geq 13\,000}$\,K. However, generally, despite this exceptional example, the discrepancy between FeSi and scaled model atmospheres can be as much as 2--3\% though it is still obvious that they are quite similar in shape.

Finally, we consider how the changes introduced by Si depend on the Fe level, i.e. the sensitivity of the Fe-peculiar models to additionally enhanced Si, in a similar way we did as for FeCr models in the previous subsection. Though, as it was mentioned above the FeSi-peculiar models are substantially different from purely Fe-peculiar models, nevertheless there is a noticeable trend that the difference between them is reduced as the Fe content grows. Overall, this trend is 1.6--2.0 times less strong than that for Fe-FeCr models (i.e. an increasing amount of Fe does not decrease the Si influence as much as it does for Cr).

\medskip\noindent\textsf{\emph{Iron--silicon--chromium}.}
The overabundance of Si, Fe and Cr simultaneously by $+1$ or $+2$\,dex gives a better overall fit to models with scaled abundances than the FeSi-peculiar models do. But for high effective temperatures (${\tef>11\,000}$\,K) additionally enhanced Si is still needed, though maybe less in value.

At this point we can conclude that the necessity of additionally overabundant Si may be attributed to the rest of the chemical elements whose influence grows with increasing effective temperature and works in a similar way as Si does. We find that among elements considered in this study, elements such as C, Mg and Ca would be good candidates to provide temperature increases in the upper atmosphere to fit temperature profiles of models with scaled abundances.

It is interesting that for the lowest effective temperature (${\tef=8000}$\,K), it is not found that Fe-peculiar model atmospheres additionally overabundant in Si or in both Si and Cr provide a much better fit to models with scaled abundances than the purely Fe-peculiar models themselves. This happens primarily because the latter already closely resemble the temperature structure of the scaled abundance models (compare them in Fig.~\ref{a-fe} for ${\tef=8000}$\,K and for the rest of $\tef$ values), and because enhanced Si and Cr result in small temperature variations for this temperature.

\medskip\noindent\textsf{\emph{Silicon--chromium}.}
Model atmospheres peculiar in both Si and Cr probably have the most interesting element combination (among Si, Cr and Fe). The reason for this follows from the fact that Si influences the model structure in a different way than Cr does, and there is no substantially dominating element like Fe. Considering the structure of the SiCr models, it is clear that Si provides its distinct temperature profile which in turn is distorted by the influence of Cr.

The main obvious effect produced by Cr is cooling of the upper atmosphere (see Fig.~\ref{cr} for purely Cr-peculiar models, as an example). If Cr is overabundant by $+2$\,dex then the temperature is decreased by about 1.5--2\% in the surface layers (${\log\tauros\leq -2}$) in comparison to Si-peculiar models, and the difference reaches 2.5\% in some cases. The Si-peculiar models enhanced in Cr by $+3$\,dex demonstrate about a 4\% decreasing in the surface layers which reaches 5--6\% sometimes. It is interesting that the cooling of the upper atmosphere due to Cr was not found to be dependent on the content of Si in any systematic way. It stays almost the same in value, varying a bit in shape regardless of the Si value.

The heating in the main line forming region provided by Cr to Si-peculiar model atmospheres is dependent on the effective temperature. For low effective temperatures (${\tef<11\,000}$\,K) its value does not depend on the Si content, most likely because Si itself does not change atmosphere structure notably for these temperatures (see Sect.~\ref{large_changes} and Fig.~\ref{si}). For higher temperatures there is a tendency that the Cr influence is being weakened with the increasing Si value. For example, the changes introduced by Cr overabundant by $+3$\,dex in the model atmosphere with the solar chemical composition (${[{\rm Si}]=0}$), on the average, are twice as large as those in the models with Si enhanced by $+3$\,dex.

\subsection{Energy distribution} \label{energy}
In this section we study the energy distribution of the calculated model atmospheres and continue the theoretical analysis of the depression at 5200\,\AA\ started in previous papers \citep[see, e.g.][]{paper2}, which is frequently observed in the spectra of CP stars.

Clearly, changes in the atmospheric temperature structure are directly related to the variations in the energy distribution. Now, after we have studied in detail the changes in the model atmosphere structure and found some regular behaviour in it, we do the same kind of analysis for the energy distribution.

We are primarily interested in the analysis of those models whose temperature distributions show the largest deviation in comparison to reference model atmospheres to understand the interplay between them. However, we also analyse the energy distributions of models with weak deviations in the temperature structure. This analysis is performed in order to insure that small changes in the temperature structure are accompanied by small changes in the energy distribution and vice versa, i.e. to support the initial element classification based on the temperature changes established in Sect.~\ref{temperature}.

Generally, the changes in the energy distribution may be local (restricted to some wavelength range) or global (for the whole wavelength range, in practice the range adopted for model atmosphere calculation). The reasons for these changes in the case of the hydrogen-rich atmospheres (our case) are discussed next.

Local changes appear as a result either of wavelength-limited spectral line absorption features (e.g. due to properties of the line density distribution of chemical elements which are important line opacity sources), or as a result of features in the continuum spectrum (discontinuities), which usually contribute essentially in a narrow UV region, and thus can be considered as local changes. These local changes are often responsible for global flux changes if located in the energetically effective regions (e.g. the far UV region for high $\tef$) because of energy redistribution: the excess or deficiency of energy in the local regions is balanced by a respective global flux deficiency or excess in the whole wavelength range. In other words, the global changes are the consequence of the local ones.

Another possibility for global-like flux changes and temperature structure changes is variation of the mean molecular weight due to different chemical compositions. However in the hydrogen-rich atmospheres the actual influence of this variation is rather small and can be neglected. We will discuss this later with respect to the extreme cases of He-weak (Sect.~\ref{small_flux}) and Si-enhanced (Sect.~\ref{large_flux}) model atmospheres.

Before we consider flux changes in detail we note that, generally, global flux redistribution due to local features grows as the effective temperature grows, because the maximum of the energy distribution shifts towards ultraviolet, increasing the amount of energy being absorbed by the continuum features or by spectral lines in the far UV region. As a result there is usually a flux deficiency in the UV and a corresponding excess in the visual region. At the same time local flux features are temperature dependant themselves, and the resulting energy redistribution behaviour is a combination of both effects.

We note also that hereafter if we describe any changes in the energy distribution without specifying explicitly its meaning we mean global changes (i.e. energy redistribution); the local changes we call flux features (e.g. the depression around 5200\,\AA, a flux deficiency or discontinuity in the UV).

The measurements of the global flux variations were performed primarily in the wavelengths interval 3000 through 7000\,\AA, i.e. in the region which is not largely congested with multiple absorption lines, unlike the far UV region, where the most local absorption occurs.

The redistributed energy generally appears as a vertical shift with respect to the reference model atmosphere fluxes. However, this shift is not uniform over a wide wavelength interval, i.e. there is a continuum inclination with respect to the reference fluxes. This inclination is a key feature of the energy distribution of CP stars that makes it different (in the visual region it mimics fluxes produced by hotter and in the UV region by cooler normal stars). The question of the inclination of the continuum, which is closely connected to the calibration of photometric indices, is outside the scope of this work. It will be considered in detail as part of a forthcoming paper.

For comparison purposes all fluxes were convolved with a gaussian with ${\rm FWHM}=15$\,\AA, and comparisons were made on a logarithmic scale ${2.5\cdot\log F_\lambda}$ (i.e. magnitudes), where $F_\lambda$ has units of ${{\rm erg\,s^{-1}\,cm^{-2}\,\AA^{-1}}}$.

Note that in the series of our papers about magnetic line blanketing \citep[see e.g.][]{paper2} we used units of ${\log F_\lambda}$ to measure flux discrepancies. The label ``magnitudes" thus has a different meaning in these earlier papers.

Below, we use the same grouping of chemical elements that we used in Sect.~\ref{temperature}.

\subsubsection{Small changes group}\label{small_flux}

The changes for model atmospheres peculiar in Sr, Eu, Hg, ${[{\rm Mg}]=-1}$, and ${[{\rm Ca}]=-1}$ are less than 2--3\,mmag. For model atmospheres peculiar in He (for any considered values) the changes are about 10\,mmag. We found that the reason for these changes (both temperature and flux changes) for low effective temperatures (${\tef\lesssim 11\,000}$\,K) is the variation (decrease by 18.5\%) of the mean molecular weight (recall that the temperature changes due to depleted He are less than 0.3\%). As we go to higher effective temperatures, both the mean molecular weight and an absorption edge of \ion{He}{i} at 504\,\AA\ are responsible for the changes in the temperature structure and flux with the growing influence of the latter one (recall again that He-depleted models produce temperature changes less than 0.5\% even for high temperatures, see Sect.~\ref{small_changes}). All the models considered in this paragraph show very small changes in the temperature structure (which are well below 1\%, see Sect.~\ref{small_changes}), and so do the fluxes they produce.

For model atmospheres with Ca overabundant by $+2$\,dex (special case model atmospheres which were assigned to the small changes group in Sect.~\ref{special_changes}), the flux changes are due to the flux discontinuity of \ion{Ca}{ii} at 1218\,\AA\ and 1420\,\AA. The maximum local discrepancy between peculiar and reference model fluxes is about 4.0\,mag deep, 100\,\AA\ wide, located bluewards from the 1218\,\AA\ feature, and appears for ${\tef=11\,000}$\,K (look at the temperature deviation from the reference model for this effective temperature, see Sect.~\ref{special_changes}). The global changes for ${\tef=11\,000}$\,K, however, are only about 15\,mmag in value. The maximum global flux variation of 25--40\,mmag appears for ${\tef=15\,000}$\,K, and for ${\tef=20\,000}$\,K, it is about 10--25\,mmag.

Model atmospheres deficient in C and CNO demonstrate fluxes which differ from the reference model fluxes less than 2--10\,mmag for low effective temperatures (${\tef\leq 11\,000}$\,K) and reach values of 10--25\,mmag for higher temperatures. The changes are due to flux discontinuities (absorption edges) in the UV region due to \ion{C}{i}, at 1101\,\AA\ and 1240\,\AA.

\subsubsection{Moderate changes group}\label{moderate_flux}

The changes in the flux distribution for model atmospheres peculiar in Ti are less than 3\,mmag. We suppose that additional absorption in spectral lines of Ti in the UV is responsible for this small flux redistribution. Although, in the temperature structure analysis (Sect.~\ref{moderate_changes}), Ti was assigned to the group of elements producing moderate changes, the weak variation in the energy distribution does not clearly support this assignment. The flux changes demonstrated by Ti-peculiar model atmospheres are actually less than those produced by models peculiar in C or Ca (see the previous section). To explain this behaviour, we recall (see Sect.~\ref{moderate_changes}) that Ti-peculiar model atmospheres show the temperature deviation more than 1\% only in the upper atmosphere (${\log\tauros<-3}$) where there is little impact on the global flux distribution, since the main line forming region is well below this depth. Thus, Ti can now be assigned to the group of elements which provide small changes.

Model atmospheres with enhanced Mn demonstrate energy distribution changes due to many absorption lines in the UV region. The redistribution effect grows with increasing effective temperature. For models overabundant in Mn by $+3$\,dex, the changes in the energy are about 90--120\,mmag for high effective temperatures (${\tef>11\,000}$\,K) and are 50--70\,mmag for lower temperatures. If we consider model atmospheres with ${[{\rm Mn}]=+2}$ then these values are 40--50\,mmag and 10--30\,mmag for high and low temperatures respectively.

Model atmospheres peculiar in Ni exhibit flux changes of about 25\,mmag for low effective temperatures (${\tef\leq 11\,000}$\,K) and about 30--60\,mmag for higher temperatures. The changes are due to UV line absorption.

Models overabundant in Mg by $+2$\,dex (special case model atmospheres which were assigned to the moderate changes group in Sect.~\ref{special_changes}) show the strongest variation in the temperature structure among elements in the moderate changes group. This is a consequence of the influence of several continuum features of \ion{Mg}{i} at 2515\AA\ and \ion{Mg}{ii} at 1169\AA\ and 1943\AA. The global flux variation with respect to reference model fluxes is about 50--100\,mmag on the average. The largest effect of about 70--100\,mmag occurs at ${\tef=15\,000}$\,K. It is interesting that the model atmosphere with ${\tef=8000}$\,K shows almost the same high values of the flux change because of a 1000\,\AA\ wide (bluewards from 2515\,\AA) and 2.0\,mag deep continuum feature which, even for this low effective temperature, absorbs enough energy to produce the same effect as occurs in a model with almost twice the temperature.

\subsubsection{Large changes group}\label{large_flux}

\medskip\noindent\textsf{\emph{Silicon}.}
Model atmospheres with enhanced Si show substantial variation of the flux in their spectra. For ${[{\rm Si}]=+1}$ these changes are about 10--40\,mmag for low effective temperatures (${\tef\leq 11\,000}$\,K) and about 25--40\,mmag for higher temperature. For ${[{\rm Si}]=+2}$, these values are 60--170\,mmag and 80--170\,mmag. Finally, for ${[{\rm Si}]=+3}$ they are about 100--250\,mmag and 200--400\,mmag, respectively.

The reason for these changes is several large flux discontinuities due to \ion{Si}{i} in the UV region (such as a feature at 1526\,\AA\ due to ionization from the ground state, or a feature at 1682\,\AA\ due to ionization from the first exited state) and strong absorption features in the UV region from $\sim$1100 to $\sim$1600\,\AA\ due to \ion{Si}{ii}. We also found that there is a weak influence on the temperature structure and fluxes due to the variation of the mean molecular weight for Si-peculiar model atmospheres overabundant in Si by $+3$\,dex. The content of Si (0.0295) is comparable with the content of He (0.0483) for these model atmospheres and the mean molecular weight increases by $+56$\% with respect to the solar composition model. However, we estimate that the resulting effect on the temperature structure due to this variation is less than 0.10--0.15\% in the main line forming region and less than 0.6\% for the upper atmosphere (${\log\tauros<-4}$) in comparison to the reference models. In fluxes, molecular weight variations can be responsible for 15--30\,mmag changes for the lowest and about 5--10\,mmag changes for the highest effective temperature. Despite the fact that the increase of the mean molecular weight is rather large, the main effect on the model atmosphere structure appears as a variation of the atmospheric geometrical size -- namely, the atmosphere contracts. However, as we mentioned above, there are only quite small changes in the optical depth scale.

Here we recall the interesting feature found in the temperature distribution for the model atmosphere with ${\tef=8000}$\,K and peculiar in Si by $+3$\,dex (see Fig.~\ref{si}). We find that the reason for this extraordinary temperature behaviour is that there are strong line absorptions and a number of distinctive flux discontinuities due to Si in the visual region ($\sim$4000--7000\,\AA), which provide a unique broadband influence that is about 0.5\,mag deep with respect to the hydrogen continuum before the temperature correction procedure (in this case flux discontinuities give an example of non-local changes). These flux features are quickly weakened with increasing effective temperature and are almost absent for the next grid effective temperature value of 9500\,K.


There is also an important contribution due to Si spectral lines in the flux depression around 5200\,\AA\ for low effective temperatures. For instance, for Si overabundant by $+1$\,dex, there is a noticeable absorption contribution in the 5200--5500\,\AA\ region for ${\tef=8000}$\,K. If Si enhanced by $+2$\,dex then this contribution is visible up to 11\,000\,K in the same wavelength range. For a value of $+3$\,dex the line absorption in the 5200--5500\,\AA\ region is clearly visible for effective temperatures up to ${\tef=13\,000}$\,K. The reason for this feature is several strong \ion{Si}{i} multiplets concentrated near 5265, 5340, 5415 and 5495\,\AA.

\medskip\noindent\textsf{\emph{Chromium}.}
For Cr-peculiar model atmospheres, if Cr is overabundant just by $+1$\,dex, the flux changes are less than 10\,mmag for low effective temperatures (${\tef\leq 11\,000}$\,K), and less than 25\,mmag for higher effective temperatures. In case of ${[{\rm Cr}]=+2}$ the changes are 40--50\,mmag and 40--70\,mmag, respectively. Finally, if Cr is enhanced by $+3$\,dex, then changes are 40--120\,mmag for low effective temperatures and 100--150\,mmag for high ones. 

The energy redistribution happens due to many absorption lines of Cr in the ultraviolet region which effectively absorb energy for the whole range of effective temperatures.

The main contribution to the 5200\,\AA\ depression is because of the absorption in the region of 5150--5450\,\AA\ due to numerous spectral lines of Cr. The effect grows with the growing Cr abundance and decreases with increasing effective temperature. The contribution can be seen clearly up to ${\tef=11\,000}$\,K if the highest overabundance value of ${[{\rm Cr}]=+3}$ is adopted.

\medskip\noindent\textsf{\emph{Iron}.}
Model atmospheres peculiar in Fe by $+1$\,dex show flux changes of about 50\,mmag for low effective temperatures (${\tef\leq 11\,000}$\,K) and 70--120\,mmag for higher effective temperatures. For models with Fe overabundant by $+2$\,dex, these values are 120--170\,mmag and 200--370\,mmag respectively.

There are several flux discontinuities in the UV region due to Fe, but they only appear for the lowest effective temperature, and thus do not contribute substantially to the energy redistribution. The global flux changes are primarily because of energy absorption in the UV region by a great number of Fe lines for all effective temperatures. We found that the influence of the altered value of the mean molecular weight ($+11$\% for ${[{\rm Fe}]=+2}$) is absolutely negligible.

Also, Fe contributes significantly to the 5200\,\AA\ depression due to a great number of Fe spectral lines absorbing energy in a wide wavelength range from 5000 to 5500\,\AA. This contribution has a weak tendency to be stronger a bit bluewards of 5200\,\AA. If Fe is enhanced by $+1$\,dex then the depression is seen for effective temperatures up to 15\,000\,K. If Fe is overabundant by $+2$\,dex then the depression is clearly recognizable up to the highest effective temperature of 20\,000\,K, gradually decreasing with increasing effective temperature.

\medskip\noindent\textsf{\emph{Scaled abundances}.}
Model atmospheres with a chemical composition scaled by $+1$\,dex produce fluxes which differ from the reference model atmosphere fluxes by 70--200\,mmag for low effective temperatures (${\tef\leq 11\,000}$\,K) and by about 120--250\,mmag for higher temperatures. For model atmospheres with ${\rm [M/H]=+2}$, flux changes are about 120--400\,mmag for low effective temperatures (${\tef\leq 11\,000}$\,K) and about 350--500\,mmag for higher effective temperatures.

Concerning the flux depression around 5200\,\AA, there is clearly a combined contribution due to Si, Cr and Fe to this depression. Putting together their influences as outlined above, we conclude that Fe is the principal contributor for the whole range of effective temperatures, while Cr and Si are important primarily for low effective temperatures.

\subsection{Photometry (the $\uvbyb$ and $\Delta a$ systems)} \label{photometry}

Photometry provides us with a powerful and well-developed tool for a preliminary analysis of observed stellar fluxes, capable of rapid measurement of a large number of stars. Quite often the photometric indices are the only available data about the stellar energy distribution, which makes the theoretical photometry analysis especially valuable.

Various photometric systems have been designed to work in different spectral intervals to probe specific broad- or narrow-band spectral features. In this section we consider two photometric systems. The first one is the Str\"omgren-Crawford $\uvbyb$ photometric system which is a general one. The second system is the peculiarity index $a$ designed to measure the strength of the 5200\,\AA\ depression with respect to normal stars (the $\Delta a$ photometric system).

We computed synthetic photometric colors in these two systems following the procedure used in our previous studies \citep[e.g.][]{paper2,paper3}. There is also a detailed description of how to calculate the synthetic $a$ index by \citet[][Sect.~2.1]{kupka1}.

An in-depth analysis of the photometric indices is beyond the scope of this paper, thus here we restrict ourselves to a summary of the photometric changes found. First we present a general discussion of the photometric index values for various models, than we describe the $a$ vs. $b-y$ diagram specifically for model atmospheres of the large changes group. All values were rounded to 1\,mmag accuracy.

A thorough investigation of the spectral energy distribution, photometric colors and calibration of photometric systems will be presented in a forthcoming paper.

\subsubsection{Discussion on the photometric indices}

The results in this section are presented with respect to the reference model atmospheres.

\medskip\noindent\textsf{\emph{Small changes group}.}
On the whole, the variations of all photometric indices of the $\uvbyb$ system in this group are approximately within a $\pm 3$\,mmag limit. However there are quite a number of exceptions to this rule which can be rather large, primarily for $m_1$ for the lowest effective temperature (8000\,K) and for $c_1$ for all temperatures. The reasons for this are various line absorption features (such as H and K \ion{Ca}{ii} lines or \ion{Sr}{ii} and \ion{Eu}{ii} lines) and the global energy redistribution. An important thing to notice is that none of the models produce changes in ${b-y}$ of more than $\pm 2$\,mmag, except those with ${[{\rm Ti}]=+2}$ for low effective temperatures (e.g. ${\Delta(b-y)=+13}$\,mmag for ${\tef=8000}$\,K). Finally, no models show variations in the peculiarity index $a$ of more than $+2$\,mmag.

\medskip\noindent\textsf{\emph{Moderate changes group}.}
It is hard to point to any particular local features contributing to a certain photometric filter which cause changes to the derived photometric indices for models in this group. We suppose that the main cause of the variations is the global energy redistribution which affects values obtained in all photometric filters. The indices of the $\uvbyb$ system show positive as well as negative variations, while the most extreme values are for low effective temperatures. The temperature indicator ${b-y}$ exhibits deviations of about $\pm 4$\,mmag with some exceptions (e.g. mean deviation of about $-15$\,mmag for models with ${[{\rm Mn}]=+3}$). The variations for $a$ are about $\pm 5$\,mmag showing extreme values of $\pm 15$\,mmag for the lowest $\tef$.

\medskip\noindent\textsf{\emph{Large changes group (Si, Fe, Cr and [M/H])}.}
The global energy redistribution and local flux depressions are especially strong for this group of elements and are responsible for changes in all indices.

All the models show positive changes in the $m_1$ index with respect to the reference model atmospheres, with some exceptions (up to $-18$\,mmag) for purely Cr-peculiar models (${[{\rm Cr}]\leq +3}$). The model atmospheres which are peculiar in one or several elements by no more than $+1$\,dex show typically about 5--20\,mmag variations which can rise to 20--50\,mmag (primarily for ${\tef=8000}$\,K). If one or several elements are overabundant by $+2$\,dex then these changes are usually about 10--100\,mmag (with largest values of 30--170\,mmag). Finally, if one or several elements are enhanced by $+3$\,dex then changes are about 20--120\,mmag, with maximum values of 40--180\,mmag. Note that the smallest $\Delta m_1$ values are for the purely Cr-peculiar models.

Generally, all the models demonstrate negative variations in the $c_1$ index (i.e. the Balmer jump becomes smaller), but there are some exceptions for low effective temperatures (${\tef\leq 9500}$\,K) showing positive values for Si- and Cr-peculiar models (up to $+32$\,mmag). The model atmospheres which are peculiar in one or several elements by no more than $+1$\,dex show about $-5$ to $-50$\,mmag variations, which peak to $-10$ to $-70$\,mmag. If one or several elements are overabundant by $+2$\,dex then these changes are about $-15$ to $-200$\,mmag (with peaks at $-20$ to $-300$\,mmag). Finally, if one or several elements are enhanced by $+3$\,dex then changes are about $-30$ to $-200$\,mmag with peaks at $-40$ to $-450$\,mmag. Note that the highest $\Delta c_1$ values are for the purely Cr-peculiar models.

For the $\beta$ index, positive $\Delta\beta$ values tend to occur at the low end of the effective temperature scale while negative values occur at its high end. The majority of the values are concentrated within a $\pm 10$\,mmag range and peak up to about $\pm 25$\,mmag for models strongly peculiar in Si for low effective temperatures.

The ${b-y}$ and $a$ indices are considered in the next section.

\subsubsection{The diagram of the peculiarity index $a$ vs. $b-y$}

For the second part of the photometry analysis we consider the diagram of $a$ vs. ${b-y}$. Thus we can explore the $\Delta a$ photometric system the way it was designed, i.e. with respect to the normal stars of the same temperature indicator color (the ${b-y}$ index in our case).

\begin{figure}
\filps{\hsize}{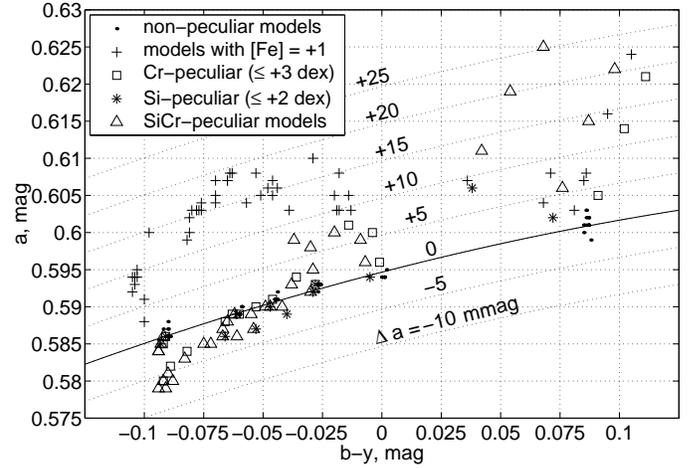}
\caption{Diagram of the peculiarity index $a$ vs. the effective temperature indicator ${b-y}$. The figure represents model atmospheres of the large changes group (except models with strong abundances of ${[{\rm Si}]=+3}$ and ${[{\rm Fe}]=+2}$). The non-peculiar models are model atmospheres assigned to the small changes group (Sect.~\ref{small_changes}), and indicated on the plot with dots ($\bullet$). All models which are peculiar in Fe by $+1$\,dex are shown with a plus sign ($+$), i.e. this includes purely Fe-peculiar, scaled, FeCr-, FeSi- and FeSiCr-peculiar model atmospheres. Models peculiar in Cr and Si only are shown with the square ($\square$) and the asterisk ($\ast$) symbols, respectively. The upward pointing triangles ($\triangle$) represent SiCr-peculiar model atmospheres. The black solid line is the normality line $a_0$ of the peculiarity index ${\Delta a=a-a_0}$, and is a parabolic least squares fit through points which correspond to the reference model atmospheres. The inclined dotted lines represent isolines of the $\Delta a$ index spaced 5\,mmag.}
\label{bya}
\end{figure}

Figure~\ref{bya} represents model atmospheres of the large changes group computed in this work (except those with strong overabundance values of ${[{\rm Si}]=+3}$ and ${[{\rm Fe}]=+2}$, to decrease number of points in the plot and provide better picture resolution and thus visualization). Note that the conclusions drawn below (illustrated on the basis of the limited sample in Fig.~\ref{bya}) are for the whole parameter space. We advise the reader to read thoroughly the caption of the figure for details. The analysis revealed several interesting features:

\begin{enumerate}
\item An important feature to notice is that none of the models peculiar in Fe (the plus (+) symbols) show negative $\Delta a$ values, they all appear above the normality line.
\item Model atmospheres which show negative $\Delta a$ values have ${\tef\geq 11\,000}$\,K and are primarily concentrated leftwards from ${b-y=-0.045}$ (the value shown by the reference model atmosphere of ${\tef=13\,000}$\,K). All of these models have the solar content of Fe.
\item The locus of points in the right upper section shows the regular highly organized pattern which we investigate below. The same feature is also clearly recognizable in the middle of the figure leftwards from the vertical zero line.
\end{enumerate}

\begin{figure}
\figps{\hsize}{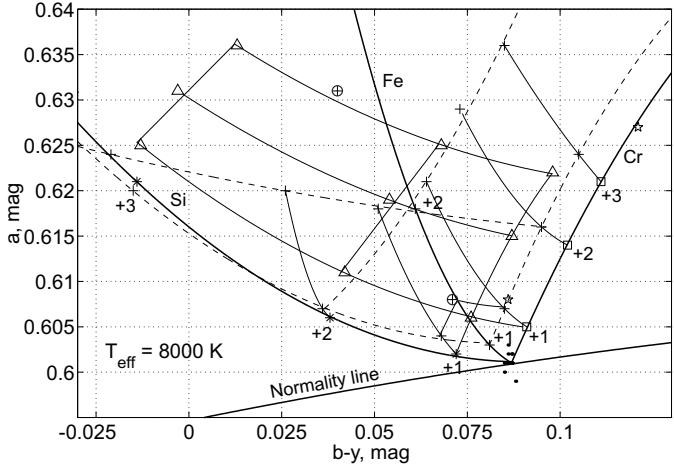}
\caption{Diagram of the peculiarity index $a$ vs. ${b-y}$ for all model atmospheres of the large and small changes groups with ${\tef=8000}$\,K. Models are represented with the same symbols as in Fig.~\ref{bya}. The plus sign ($+$) retained for all Fe-peculiar models, while the crossed circle ($\oplus$) is for SiCrFe models and the star symbol ($\star$) is for models with scaled abundances. Thick lines with numerical labels in dex units represent peculiarity axes of Si, Cr and Fe, e.g. axes formed by model atmospheres peculiar only in one of these elements. For visualization purposes there are isolines on the plot: dashed lines are the isolines of the Fe-axis, the rest of isolines are the solid lines. All curves are least squares parabolic or linear fits through the corresponding points.}
\label{bya_08000_c}
\end{figure}

\begin{figure}
\figps{\hsize}{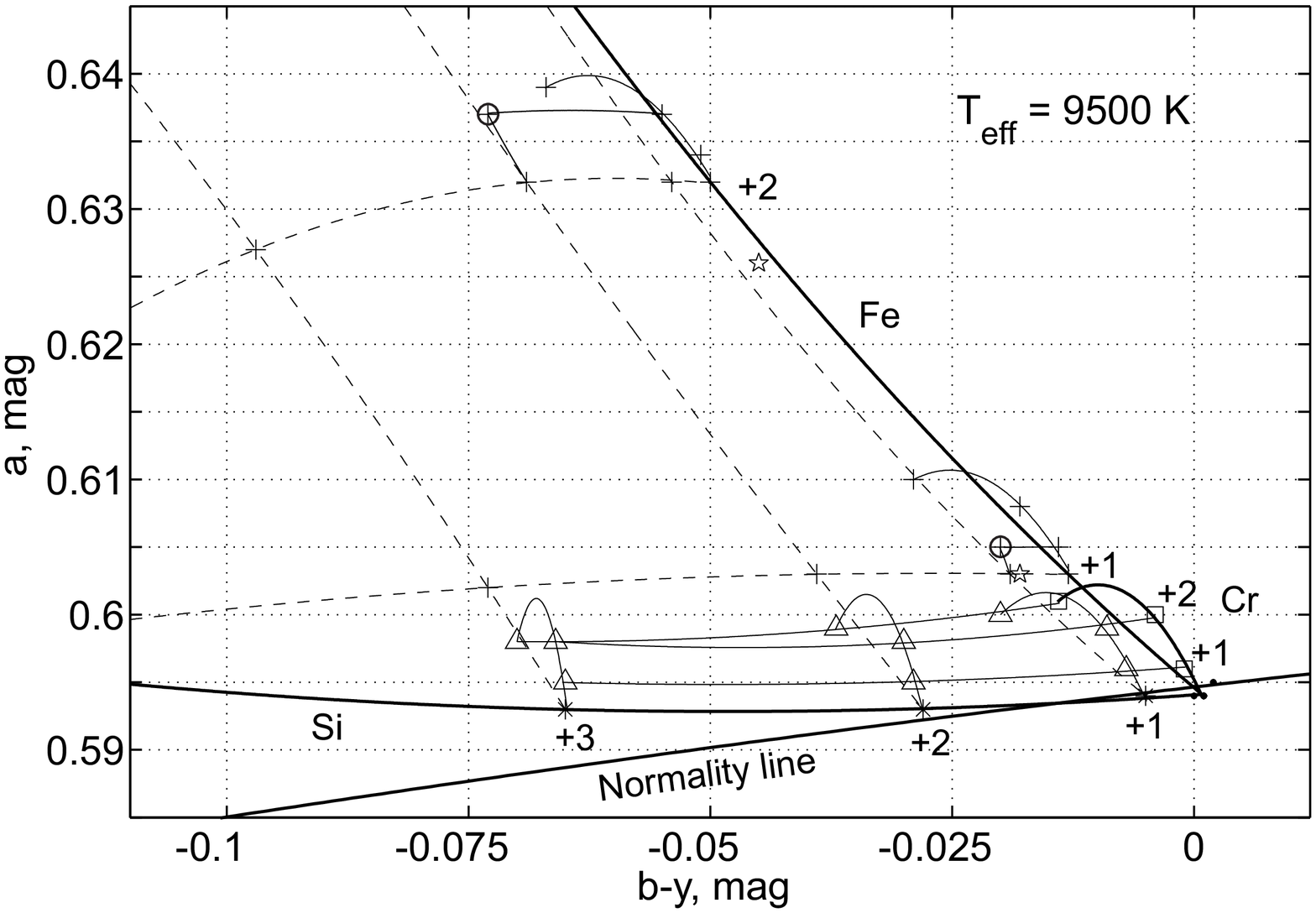}
\caption{The same diagram as in Fig.~\ref{bya_08000_c} but for ${\tef=9500}$\,K. Dash lines in this plot are the isolines of the Si-Fe grid. Note that parabolic fits to the indices of the SiCr models (see connected triangles $\triangle$) may not be a fully adequate approximation in this particular case.}
\label{bya_09500_c}
\end{figure}

\begin{figure}
\figps{\hsize}{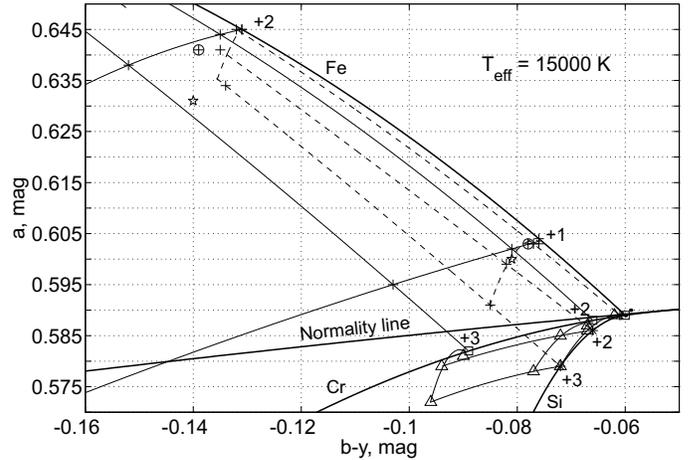}
\caption{The same diagram as in Fig.~\ref{bya_08000_c} but for ${\tef=15\,000}$\,K. Dash lines in this plot are the isolines of the Si-Fe grid.}
\label{bya_15000_c}
\end{figure}

To understand the regular pattern found in Fig.~\ref{bya} we plotted three separate ${b-y}$ vs. $a$ diagrams for models of different effective temperatures (Figs.~\ref{bya_08000_c}--\ref{bya_15000_c}). We have chosen values of 8000\,K, 9500\,K and 15\,000\,K because the reference model atmospheres of these effective temperatures on the ${b-y}$ scale represent the left, middle and the right parts of the diagram shown in Fig.~\ref{bya}. Thus, this way we investigate two patterns found above (point number 3) and investigate whether there is the same regular behaviour for the high temperature models which is harder to identify within the diagram in Fig.~\ref{bya}.

Figure~\ref{bya_08000_c} shows the ${b-y}$ vs. $a$ diagram for ${\tef=8000}$\,K. The caption for the figure provides the explicit description of symbols and labels. The figure demonstrates that there are three main directions (or axes) on the diagram associated with growing content of each of the elements Si, Cr and Fe. Moreover, there is an obvious hint that all the models tend to be located in the intersections of the isolines. For example, if we sequentially follow along the Si, Cr and Fe axes by the amount corresponding to a $+1$\,dex change in each move then we get to the point where the model atmosphere overabundant in Si, Cr and Fe by $+1$\,dex is located (the one marked with the symbol $\oplus$ in the right bottom part of the diagram). Consider another example, of the model atmosphere overabundant in Si by $+2$\,dex and in Cr by $+3$\,dex. Again, to get to the point on the diagram which represents this model we should follow the Si direction to the point labelled $+2$\,dex and then along the Cr direction to the point with the $+3$\,dex mark.

Of course, there are not perfectly concordantly aligned isolines everywhere on the diagram. For instance, in the left part of the diagram, there is a pattern break for model atmospheres peculiar in Si by $+3$\,dex. The distance between the Si-axis marks stretches slower than the corresponding one between nodes on the isolines of the SiCr-peculiar models (see connected triangles ($\triangle$) symbols along the Si direction on the diagram).

Similar diagrams for ${\tef=9500}$ and 15\,000\,K are shown in Figs.~\ref{bya_09500_c} and \ref{bya_15000_c}, respectively. The analysis of all these figures and those for other effective temperatures reveals some interesting properties:

\begin{enumerate}
\item As the effective temperature grows the low curvature Si-axis gradually rotates counter-clockwise sinking below the normality line at around 10\,000\,K.
\item The Cr-axis shows the same behavior, going below the normality line at around 13\,000\,K. At the same time it changes its shape from the initially straight line to the arched one.
\item Finally, in contrast to items above, the Fe-axis is very stable in its shape and inclination. In fact, it is almost straight and virtually perpendicular (on the scale of Fig.~\ref{bya_08000_c}) to the normality line for any effective temperature. If all the Fe axes for different effective temperatures are plotted on the same diagram then they appear to be almost perfectly parallel.
\end{enumerate}

This analysis supports our earlier conclusion that Fe is the main contributor into the 5200\,\AA\ depression for the whole range of effective temperatures (Sect.~\ref{large_flux}).

It also demonstrates that Fe is indeed the key element for the fundamental property of the $\Delta a$ system to show positive values for CP stars with overabundant Fe-peak elements in their atmospheres, and negative values for CP stars with underabundances of the Fe-peak elements ($\lambda$~Bootis stars) \citep[see the summary on $\Delta a$ measurements by][Table~3]{paunzen2005}.

The diagram analysis confirmed the role of Cr and Si as important contributors into the magnitude of the 5200\,\AA\ depression (measured through the $\Delta a$ system) for low effective temperatures.

To extend our study and ensure the reliability of the diagram properties outlined above we computed several additional model atmospheres for the effective temperature of 7000\,K and found indeed for them the same features.

Also we found that model atmospheres scaled by $+1$\,dex show ${b-y}$ values which differ from those of the reference model atmospheres by a small to modest amounts: $-1$, $-19$, $-24$, $-25$, $-21$, $-10$\,mmag (in order of increasing effective temperature). These values are actually in a perfect agreement with numerical results by \citet{kurucz13}, and their modest values partly explain why we did not find any systematic errors in the determination of the fundamental parameters of magnetic CP stars with scaled abundances in their atmospheres \citep{paper2}.

At the same time Figs.~\ref{bya_08000_c}--\ref{bya_15000_c} show that if the model atmosphere is overabundant in Si, Cr or Fe (separately or together) by more than $+1$\,dex then this may result in a quite large deviation of the temperature indicator ${b-y}$, that is explainable by the flux redistribution effects (see Sect.~\ref{large_flux}). However, in some cases simultaneous overabundance of two or even three of these elements produces less variation of the ${b-y}$ value than they would separately.

\subsection{Hydrogen line profiles}

\begin{figure*}
\filps{\hsize}{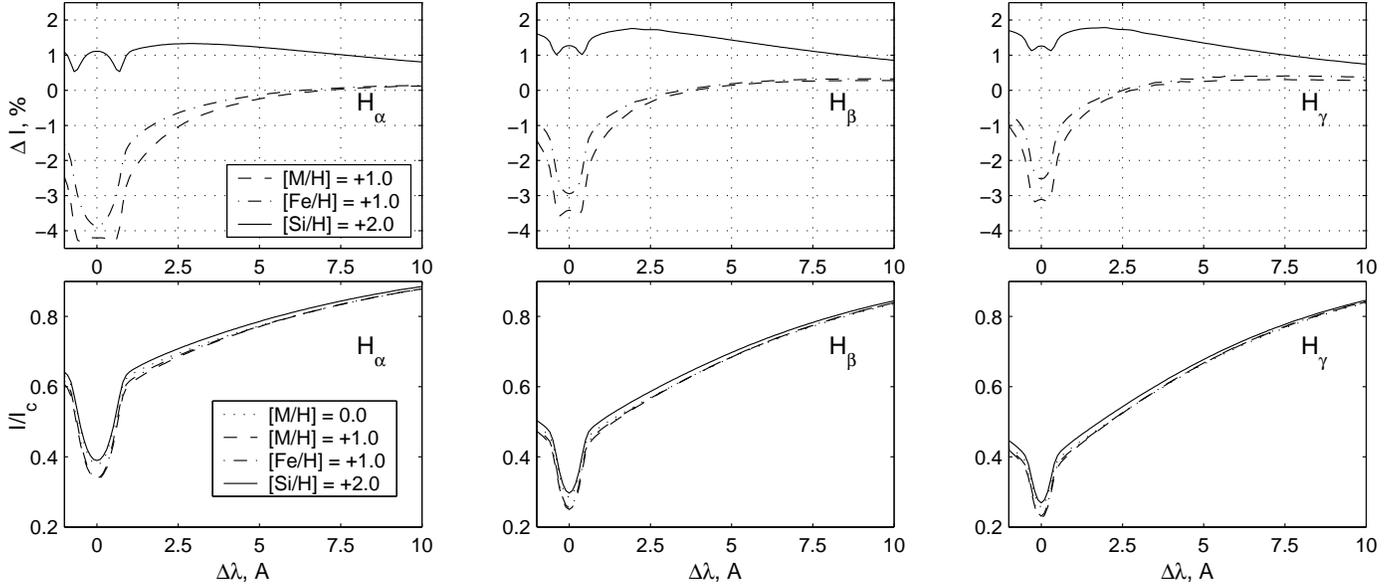}
\caption{Comparison between H$\alpha$, H$\beta$ and H$\gamma$ line profiles computed for model atmospheres with ${\tef=13\,000}$\,K and different chemical composition (peculiar abundances ${\rm [M/H]=+1}$, ${[{\rm Fe}]=+1}$, ${[{\rm Si}]=+2}$ and the reference solar abundances ${\rm [M/H]=0}$). The upper subplots show the difference (${\Delta I=I-I_{\rm [M/H]=0}}$) between line profiles calculated for models with peculiar abundances and those for the reference model atmospheres. The normalized line profiles (${I/I_c}$) are represented in the bottom subplots.}
\label{balmer}
\end{figure*}

To study the effects due to abundances other than solar on the hydrogen line profiles we calculated profiles of the H$\alpha$, H$\beta$ and H$\gamma$ lines and compared them with those for reference model atmospheres.

In accordance with the results obtained in Sect.~\ref{temperature} (model atmosphere temperature structure) we decided that model atmospheres with peculiar abundances such as ${\rm [M/H]=+1}$, ${[{\rm Fe}]=+1}$ and ${[{\rm Si}]=+2}$ would be sufficiently representative for the analysis of the hydrogen line profiles. The comparison between models with an effective temperature of 13\,000\,K, which we find to be the best case to illustrate the following discussion, is represented in Fig.~\ref{balmer}.

We found that hydrogen line profiles for models with ${\rm [M/H]=+1}$ and ${[{\rm Fe}]=+1}$ are essentially the same for high effective temperatures (${\tef\geq 11\,000}$\,K) and show a noticeable negative deviation from the reference line profiles in the region about $\pm 5$\,\AA\ around the line center. Moreover, the difference between peculiar and reference line profiles almost does not change as the effective temperature changes. For low effective temperatures (${\tef\leq 9500}$\,K), the line profiles for the peculiar models do show broad negative variations (especially for 8000\,K) from the reference line profiles and differ from each other by about 1--2\% (models with scaled abundances produce a larger effect). Overall we find that the maximum variation in profiles of H$\alpha$, H$\beta$ or H$\gamma$ is about $-4$\% (generally, the strongest variation is for H$\alpha$, the weakest for H$\gamma$).

Model atmospheres peculiar in Si show primarily negative and broad line profile variations for low effective temperatures. These changes are about $-1$\% to $-2$\% for models with ${\tef=8000}$\,K and about $-3$\% for ${\tef=9500}$\,K. For higher effective temperatures the differences between line profiles of peculiar and reference model atmospheres are still broad but they become positive showing variations of up to $\sim+2$\%.

\subsection{Abundance analysis}

In this section we present results of a brief investigation on abundance analysis in the atmospheres of CP stars. An introduction to the problem is given in Sect.~\ref{introduction2}. We study the effects of the individual abundance patterns on the abundance determination procedure through the spectrum synthesis method. The idea of the method is to fit theoretical calculations to the observations. Usually, this involves the initial choice of the model atmosphere (generally calculated for a scaled solar chemical composition) which is believed to be the best approximation, and the following use of a spectrum synthesis code to compute the high resolution spectrum. The atmospheric abundances are to be varied during the synthesis procedure to match the synthetic spectrum to the observed one.

The questionable point of this method is that the model atmosphere which is used for the spectrum synthesis is considered as a parameter, not as a function. In other words, the model is not recalculated in accordance with the results of the abundance analysis.

We study the effects of this approach as applied to the abundance analysis of Si, Cr and Fe (elements of the large changes group, Sect.~\ref{large_changes}). The study covers abundance values of Si and Fe up to $+1$\,dex and Cr up to $+2$\,dex.

First, we calculate a reference spectrum for a particular model atmosphere from our grid (Si-, Cr- and/or Fe-peculiar model). Second, we use model atmospheres for scaled chemical compositions (${\rm [M/H]=0}$ and ${\rm [M/H]=+1}$) to synthesize two fitting spectra, assuming the same abundances of Si, Cr or Fe that were used to compute the reference spectrum. Then, we compare each of the fitting spectra to the reference one and, if necessary, vary the abundance values of Si, Cr or Fe in order to fit the reference spectrum. The changes in abundances required to fit the reference spectrum reflect errors or other unwanted, possibly systematic, effects following from the use of the incorrect model atmosphere. All deviations are measured with respect to the reference spectrum.

The spectral line data for the spectrum synthesis procedure was extracted from VALD. We used the same lines of Si, Cr and Fe which were adopted for the stratification analysis of HD\,204411 by \citet[][Table~2]{ryabchik_2005}. The advantage of this compilation is that all the lines are carefully selected, representing samples of the low- and high-excitation lines of neutral and once ionized atoms. The data cover the spectral range from $\sim$5000\,\AA\ to $\sim$6400\,\AA\ and consists of 23 lines of Cr, 23 lines of Fe and 9 lines of Si. All spectra were computed with zero microturbulence and no rotation.

Here we note that in our analysis we suppose that the hypothetical analyst knows precisely the fundamental parameters of the stars being analyzed, although this is not correct for real stars, and especially not for CP stars. There are usually also other uncertainties and difficulties in working with real spectra which influence the actual abundance analysis procedure. Our theoretical study is not intended to simulate accurately the problem the observer faces, but to answer the question of what are the ambiguities of the abundance analysis (including the analysis of vertical stratification) using classic model atmospheres, if this analysis is performed in the ideal conditions of other variables being fixed and known. For the same reason of providing homogeneity to the study, we use the same spectral lines for all effective temperatures.

\subsubsection{Fe abundance}

We use the Fe-peculiar model atmosphere with ${[{\rm Fe}]=+1}$ to calculate the reference spectrum. The comparison of the fitting spectrum (${\rm [M/H]=0}$) with the reference one showed monochromatic flux discrepancies of about $\pm 2$\% on average, with positive ($+4$\%) and negative ($-3$\%) peaks. Despite these small values we were able to find some clear systematic effects, which actually appear to be common properties for other spectra and other elements tested within this abundance analysis. Thus we describe these effects and give some examples as they apply to this particular fit:

\begin{enumerate}
\item The spectral lines of neutral and once ionized atoms show different deviations with respect to the reference spectrum, requiring different abundance corrections to fit them.

For example, for ${\tef=11\,000}$\,K, the \ion{Fe}{i} lines show negative while the \ion{Fe}{ii} lines positive variations that require negative and positive corrections to the Fe abundance, respectively.

\item It is impossible to fit the core and the wings of strong lines simultaneously with the same abundance value. This effect decreases as the effective temperature grows (and the number of strong lines goes down).

For example, for ${\tef=8000}$\,K, all the \ion{Fe}{i} lines show this effect, demonstrating more than the $\pm 3$\% deviations in the core and wings.

\item The low- and high-excitation lines and the lines of different intensities require different abundance corrections.

For example, for ${\tef=15\,000}$\,K, the high-excitation \ion{Fe}{ii} lines require an abundance value which is $\sim$0.1\,dex higher than that for low-excitation lines.

\end{enumerate}

The comparison of another fitting spectrum (${\rm [M/H]=+1}$) to the reference one demonstrates quite similar behaviour except that all changes are of the opposite sign.

Since we do not expect that only Fe would be overabundant in the real atmosphere of a CP star, we calculated a second reference spectrum using a model atmosphere peculiar in Si, Cr and Fe by $+1$\,dex. The comparison of the fitting spectrum (${\rm [M/H]=0}$) to the new reference one shows $\pm 3.5$\% monochromatic flux variations on average with positive ($+5$\%) and negative ($-6$\%) peaks along with the same properties found above.

Another fitting spectrum (${\rm [M/H]=+1}$) shows an almost perfect agreement (about 0.5\%) with the new reference spectrum in the range of effective temperatures from 9500\,K to 13\,000\,K due to a good agreement between the temperature structures of these two models (see Sect.~\ref{changes_discussion}). For other $\tef$ values the deviation is about 1--2\%.

Finally, we find that the ambiguity in the determination of the Fe abundance due to use of various lines is within a $\pm 0.25$\,dex error bar, and apparently depends on the differences in the atmosphere structure (i.e. density, pressure, temperature and other quantities influencing line profiles) of the model being tested.

\subsubsection{Cr abundance}

As in the previous subsection we adopted two models peculiar in Cr to calculate the reference spectrum. One model is peculiar only in Cr by $+2$\,dex, another one is more realistic and additionally peculiar in Fe by $+1$\,dex. The latter appears to produce the best agreement in the sample (about 1--2\% discrepancies) if compared to the ${\rm [M/H]=+1}$ fitting spectrum.

During the analysis we revealed all the same systematic effects outlined for Fe. The uncertainties found in the analysis are also about the same values as for Fe, i.e. inside a $\pm 0.25$\,dex error bar.

\subsubsection{Si abundance}

Again, we used two model atmospheres to compute the reference spectrum. The first one is the Si-peculiar model (${[{\rm Si}]=+1}$), the second one is the SiCrFe-peculiar model with all these elements overabundant by $+1$\,dex, i.e. the same model we used for the abundance analysis of Fe. The best agreement (about 1\% deviation) is found for the ${\rm [M/H]=0}$ fitting spectrum and the reference spectrum of the Si-peculiar model, and for the ${\rm [M/H]=+1}$ fitting spectrum and the reference spectrum of the SiCrFe-peculiar model. Overall, the same systematic effects and approximately the same error bar were found for Si as for Fe and Cr.

\subsubsection{Summary on the abundance analysis}

In spite of the fact that the typical error bar found for abundance analysis in the visual region is about $\pm 0.25$\,dex, the particular error value depends on the model atmosphere being used for the analysis, and varies for different effective temperatures. In fact, some models provide a very good fit to the reference spectrum with almost no error, while others result in differences beyond this error bar. Apparently, these discrepancies follow from the differences in the temperature structure of the model used for analysis and the actual atmosphere (see Sect.~\ref{changes_discussion}).

Also, considering the results of our numerical experiments as applied to stratification analysis using homogeneous model atmospheres \citep[see for instance][]{ryabchik_2005}, we find that the uncertainty of the value of the vertical abundance gradient is within an 0.4\,dex error bar.

\section{Conclusions} \label{conclusions}

We have computed and analyzed a grid of model atmospheres of chemically peculiar (CP) stars. The main purpose was to perform a systematic homogeneous analysis of the effects of the individual abundance patterns on the model atmosphere structure, energy distribution, photometric indices (in the $\uvbyb$ and $\Delta a$ systems), hydrogen line profiles, and on an abundance determination procedure.

The grid of model atmospheres consists of more than 300 models and represents the following types of stars: CP1 (Am stars), CP2 (Si, Cr-Sr-Eu A and B stars), CP3 (Hg-Mn, B-type stars) and CP4 (He-weak, B-type stars). All models were calculated with the \llm\ code (version 8.4) assuming no magnetic field and no convection to highlight abundance effects only. Consequently, we considered model atmospheres with different chemical composition and compare them to the reference models with solar abundances. Detailed data about the chemical elements examined and the adopted abundance values are represented in Table~\ref{abundances}.

The conclusions and main results of the study are summarized below:

\begin{enumerate}

\item The majority of the tested chemical elements (within the limits of abundance values considered) produce less than 1\% variations in the model atmosphere temperature profile and fall into the small changes group. These elements are He, C, N, O, Mg (deficient), Ca, Sr, Eu, Hg. According to the results, a 1\% limit was adopted as the error bar threshold.

\item Several elements were assigned to the moderate changes group in accordance with the effects they produce on the model atmosphere structure (1--3\%) and the energy distribution. These elements are Mn, Ni, Mg (enhanced).

\item The group of elements which produce large changes in the model atmosphere structure (more than 3\%) and energy distribution consists of Si, Cr, Fe and scaled abundance patterns.

\item If we consider changes in the temperature structure produced by the elements Si, Cr and Fe in the main line forming region then we find that the sequence of Fe-Si-Cr represents these elements from the most to the least influential.

\item Model atmospheres peculiar in Cr, Mn and Fe demonstrate a very similar, highly organized, temperature behaviour. There are two distinct cooling and heating regions in the upper and the lower (i.e. the main line forming region) atmosphere, respectively. As the abundance value grows, the temperature drops in the upper atmosphere and increases in the lower atmosphere. The inflection point which separates these two regions moves outwards with growing $\tef$. The magnitude of the heating region steadily grows with increasing $\tef$ excluding the lowest value of 8000\,K. The same effects, except for the cooling region feature, apply to models with scaled abundances as well.

The reason why Cr, Mn and Fe demonstrate such behaviour is likely the similarity of their energy level configurations (and the corresponding spectral line distribution patterns), and their relatively high content in the atmosphere.

\item We concluded that the elements Si and Fe are the main providers of two different types of the temperature changes to produce the same distinctive temperature structure that model atmospheres with scaled abundances do. In other words, the temperature structure of a model atmospheres peculiar only in Si and Fe may be close to the temperature profile of a models with scaled abundances. The agreement is quite good for low effective temperatures, while for high effective temperatures (${\tef>11\,000}$\,K) the cumulative effect of all other chemical elements is required. We suspect that overabundant C, Mg and Ca are of the most significance.

\item We confirm that model atmospheres with scaled abundance patterns can not be used to simulate accurately effects of the individual abundance patterns.

\item We found that Fe is the principal contributor into the 5200\,\AA\ depression for the whole range of effective temperatures, while Cr and Si are important primarily for low effective temperatures.

\item The analysis of the diagram of the peculiarity index $a$ vs. ${b-y}$ for the model atmospheres peculiar in Si, Cr and Fe revealed regular patterns in the locus of points representing those models. In fact, there are three separate directions (axes) associated with the growing content of each element on the diagram. The inclination of the Fe-axis to the normality line $a_0$ and the fact that it does not depend on the effective temperature  clearly demonstrate that Fe is indeed the key element to the fundamental property of the $\Delta a$ system to recognize CP stars with overabundant (${\Delta a>0}$) and underabundant (${\Delta a<0}$, i.e. for $\lambda$~Bootis stars) Fe-peak elements in their atmospheres.

\item We investigated an abundance analysis procedure based on theoretical atmospheres with individual abundance pattern using models with scaled solar composition and spectral lines in the visual region. We find that the error bar of the analysis, which occurs in the context of the ideal conditions of a theoretical study, is of order 0.25\,dex, and that the particular error value depends on the model atmosphere being used for such analysis, and varies for different effective temperatures.

\item Considering the results of our numerical experiments as they apply to the stratification analysis using homogeneous model atmospheres, we conclude that uncertainty of the value of the vertical abundance gradient is within an 0.4\,dex error bar. That, of course, does not eliminate the phenomena of the abundance stratification in the atmospheres of CP stars, however it can effectively increase or decrease the gradient value deduced from the observations. That, in turn, may be essential in comparison with the results of self-consistent diffusion calculations.

\end{enumerate}

In a forthcoming paper, we are planning to focus on the study of observed energy distributions of CP stars, their photometric indices, and the calibration procedure for the determination of fundamental parameters.

\begin{acknowledgements}
We are grateful to Dr.~J.~D.~Landstreet for useful discussions and corrections to the manuscript. This work was supported by a Postdoctoral Fellowship to S.K. at UWO funded by a Natural Science and Engineering Council of Canada Discovery Grant. S.K. and D.S. acknowledge the support by the Austrian Science Fonds (FWF-P17890).

\noindent We are also grateful to the referee for valuable comments which have led to substantial improvement in the manuscript.

\noindent S.K. appreciates multi-CPU computer facilities provided for this research by the Star Formation and Interstellar Medium group in the P\&A Department at UWO and personally to Profs.~Martin Houde and Shantanu Basu.
\end{acknowledgements}

\end{document}